 \documentclass[final,5p,times,twocolumn]{elsarticle}

\usepackage{amssymb}
\usepackage{multirow}
\usepackage{amsmath}
\usepackage{doi}

\usepackage{booktabs}
\usepackage{caption} 
\usepackage{graphicx,subfigure}
\usepackage{verbatim}
\usepackage{makecell}
\usepackage{longtable}

\usepackage{rotating}

 \usepackage{colortbl} 

\usepackage{lineno}

\begin{document}

\begin{frontmatter}

\title{NuSD: A Geant4 based simulation framework for segmented anti-neutrino detectors}


\author[label1]{Mustafa Kandemir\corref{cor1}}
\cortext[cor1]{corresponding author}
\ead{mustafa.kandemir@erdogan.edu.tr}
\author[label2,label3]{Emrah Tiras}
\author[label4]{Vincent Fischer}

\address[label1]{Department of Physics, Recep Tayyip Erdogan University, 53100, Rize, Turkey}
\address[label2]{Department of Physics, Erciyes University, 38030, Kayseri, Turkey}
\address[label3]{Department of Physics and Astronomy, The University of Iowa, 52242, Iowa City, IA, USA}
\address[label4]{Department of Physics and Astronomy, Iowa State University, 50014, Ames, IA, USA}




\begin{abstract} 
NuSD: Neutrino Segmented Detector is a Geant4-based user application that simulates inverse beta decay event in a variety of segmented scintillation detectors developed by different international collaborations. This simulation framework uses a combination of cross-programs and libraries including Geant4, ROOT and CLHEP developed and used by high energy physics community. It will enable the neutrino physics community to simulate and study neutrino interactions within different detector concepts using a single program. In addition to neutrino simulations in segmented detectors, this program can also be used for various research projects that use of scintillation detectors for different physics purposes.\\
\textbf{Program summary}  \\
\textit{Program title:} NuSD \\
\textit{CPC Library link to program files:} \\
\textit{Licensing provisions:} GNU General Public License 3 \\
\textit{Programming language:} C++ \\
\textit{External routines/libraries:} Geant4, CLHEP, ROOT, CMAKE \\
\textit{Nature of problem:} \\
There has been a quite effort on the small anti-neutrino detectors for various purposes by different collaborations around the globe, but there is not an interactive and user-friendly joint-simulation framework. \\
\textit{Solution method:} \\
To fulfill the need in the field, we developed a simulation framework to combine various segmented detector technologies. This package code will be open to public to let people simulate and test different detector concepts easily. Here, we categorized segmented anti-neutrino detectors as homogeneous and inhomogeneous detectors. Homogeneous detectors consist of PROSPECT, NULAT, and HSP type detectors and inhomogeneous detectors consist of PANDA type detectors with single scintillator modules, and CHANDLER, SOLID and SWEANY detector types with composite scintillator modules. 
 
\end{abstract}


\end{frontmatter}


\section{Introduction}
\label{sec1}

In a global context of growing energy needs and reduction of carbon emission, the use of nuclear energy as an alternative to fossil fuels is expanding and more countries desire or need to acquire the capability to generate nuclear power. 
In order to limit the possible misuse of civilian nuclear power to military purposes, the International Atomic Energy Agency (IAEA) has been looking into ways to guarantee the peaceful use of nuclear reactors through inspection and monitoring \cite{NuTool}.

Emitted in overwhelming numbers by the decay of fission products in nuclear reactors, neutrinos\footnote{In this study, electron anti-neutrinos are referred to as “neutrinos” for brevity.} have the unique property of being able to travel almost unhindered through matter, making them impossible to block or tamper with.

When emitted at low energies ($\sim$1-10~MeV), such neutrinos can be detected through the Inverse Beta Decay reaction (IBD) given its higher cross section on hydrogen~\cite{Vogel}. 
This reaction leads to the simultaneous emission of a positron and a neutron, whose detection can be correlated both in time and in space. 
A neutrino detector, placed at an optimal distance from a nuclear reactor, would thus have the capability to monitor its neutrino emission flux, directly correlated to the reactor thermal power, and its fissile content in real-time, without encountering the risk of being duped.

However, a detector placed in the vicinity of a nuclear reactor faces several challenges in terms of background and deployment. When deployed within or next to a reactor building, compactness is required in order to fit in existing galleries or rooms and possibly to be deployed in several locations in the same reactor complex. Time-correlated and accidental backgrounds from cosmic rays or the reactor itself are likely to mimic a neutrino interaction and bias the estimation of the neutrino flux. Such backgrounds can be mitigated by segmenting the detector volume into smaller sub-volumes thus allowing a better reconstruction of cosmic rays and their interaction products and a better separation of accidental backgrounds, spatially uncorrelated unlike neutrino interactions.

As such, while several projects plan on detecting low energy neutrinos through IBD using monolithic water-based detectors~\cite{WATCHMAN:2015lcq,Angra,Fischer:2020htg,Bat:2021jyq} thus allowing the reconstruction of Cherenkov patterns, segmented liquid scintillator based detectors offer a decisive advantage in terms of background reduction and detection efficiency.

Compact segmented scintillation (liquid or plastic) detectors have sparked an interest from the low energy neutrino physics and the nuclear non-proliferation communities as they are well suited to detect neutrino signals in an environment such as a nuclear reactor and its proximity. A global effort is currently ongoing to design and develop several of such detectors, each with their own strengths, weaknesses, and physics goals.

We believe that a similar effort for simulation studies of such several detectors by using a single framework would be useful. In this sense, NuSD brings together numerous reactor neutrino experiments under a single simulation environment, giving the neutrino physics community an easy-to-use, versatile and user-friendly program to explore diverse detector technologies developed by various international collaborations. 

\section{Implementations}
\label{sec2}
Geant4 \citep{Agostinelli} is an open-source C++ software package composed of tools that can be used to accurately simulate the passage of particles through matter. The toolset includes all parts of the simulation process, including \cite{G4Toolkit}:
\begin{itemize}
\setlength\itemsep{0.em}
	\item the geometry of the system and the materials involved
	\item the particles of interest
	\item the creation of primary particles
	\item the tracking of particles through materials
	\item the physics processes regulating particle interactions
	\item the response of sensitive detector components
	\item the creation of event data	
\end{itemize}

\begin{figure*}
 \centering
    \includegraphics[width=0.9\textwidth]{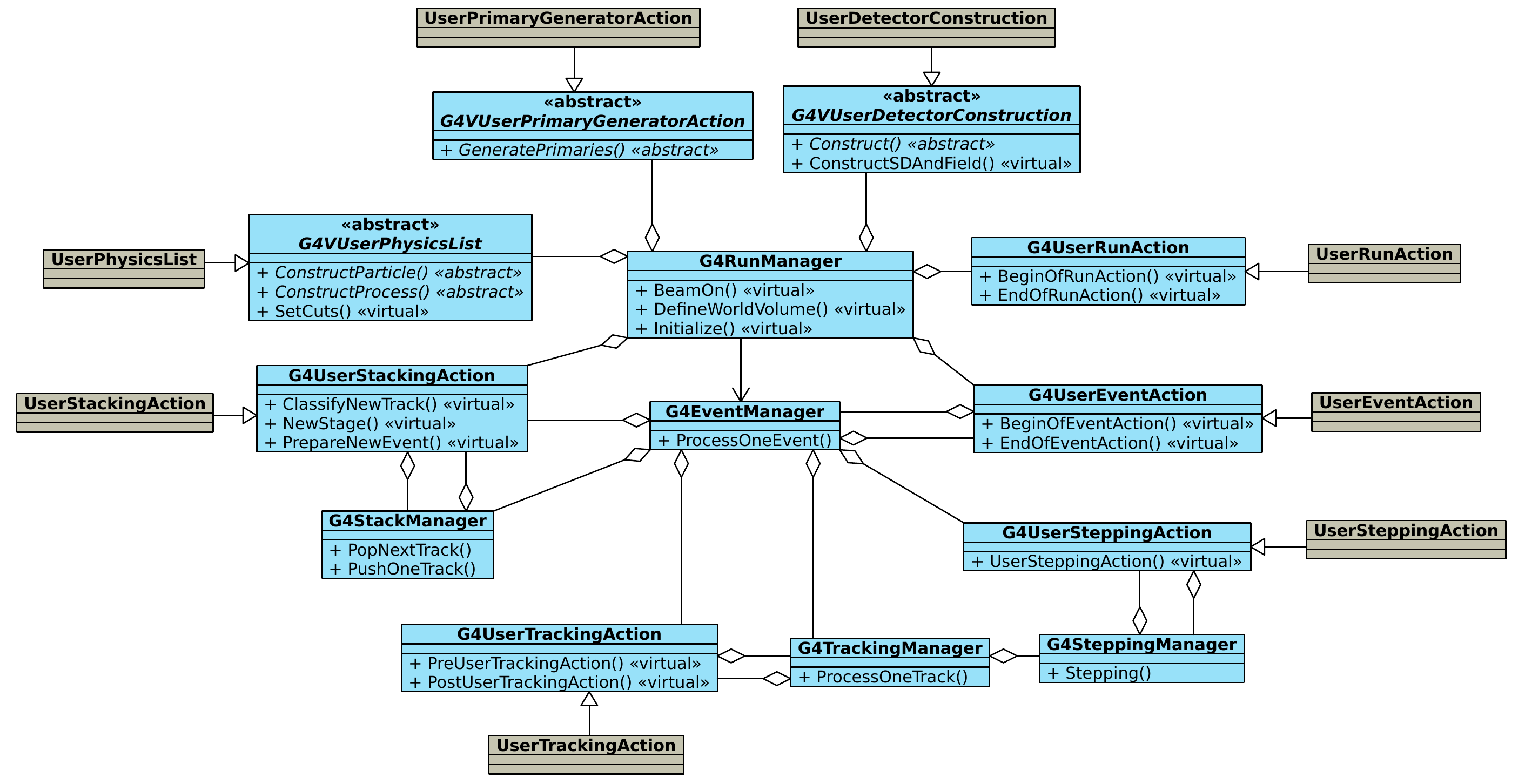}
     \caption{ Unified Modeling Language (UML) class diagram of a typical Geant4 application. The blue color shows classes provided by Geant4, and the brown color indicates concrete classes to be written by the application developers.}
    \label{fig:fig1}
\end{figure*}

Geant4 does not provide any standard program the developer or user can run. A developer must write their own C++ program based on the Geant4 user classes. Fig. \ref{fig:fig1} presents the software architecture of a typical and simple Geant4 application in the Unified Modeling Language (UML) \cite{Fowler:2015}. The group of classes marked as abstract represents classes that Geant4 developers should implement. Other classes are optional and used to get information and gain control of the simulation at various stages. The classes indicated in brown are the concrete classes to be written by the users.  

NuSD is built based on the Geant4 user classes. Apart from the option of utilizing the standard Geant4 API (Application Programming Interface) to implement user classes, NuSD builds an additional API on top of the standard Geant4 API to facilitate the implementation process. To retain flexibility, some of the user classes in NuSD are directly derived from Geant4 user classes, while others are inherited from newly created abstract classes to avoid code redundancy. During the implementation process, the inheritance concept of Object Oriented Programming (OOP) is heavily used to share behaviors between classes. The next sections go through the NuSD implementation in depth.

\subsection{Overview of NuSD}
\label{sec2.1}
NuSD is written in C++ language and is built based on Geant4 MC code. It requires ROOT \cite{Root} and Geant4 software packages to be installed on the system as a prerequisite to work. The code is compiled using CMake and tested on Linux operating system with the version of Geant4 10.06.  

NuSD utilizes a main configuration file named \textit{NuSD\textunderscore config.h} to conditionally compile certain parts of the code. To produce this file, a file named \textit{NuSD\textunderscore config.h.in} is provided in the source directory. This file contains all of the preprocessor macro definitions used throughout the program. When NuSD is built through CMake, the \textit{NuSD\textunderscore config.h.in} file is automatically converted to \textit{NuSD\textunderscore config.h} and the generated file is copied into the build directory. The users can edit this file by commenting out or activating code lines. This file includes the following preprocessor macros:

\begin{itemize}
\setlength\itemsep{0.em}
	\item  \#define DETECTOR\textunderscore NAME HSP. The user chooses one of the defined detectors. HSP is selected by default.
	\item \#define CREATE\textunderscore ROOT\textunderscore FILE. The user decides whether to output a ROOT file from the simulation.
	\item \#undef GENERIC\textunderscore PRIMARY\textunderscore GENERATOR. The user decides whether to simulate the IBD event or any other event. By default, the IBD event is simulated.
	\item \#define NuSD\textunderscore DEBUG. The user decides whether to print the results of the simulation.
		
\end{itemize}

NuSD provides many macro files for users to easily modify simulation parameters. A macro file is an ASCII file that contains user interface commands allowing easy control of the simulation and associated settings, without having to interact with the source code of the program. Thus, a NuSD user only needs basic knowledge of how to control the program flow but does not necessarily have to know object-oriented programming or C++.

NuSD works in both sequential and multi-threading modes. It is very critical to be able to run in multi-threading mode, particularly in cases where optical photons are simulated. It is because when a particle deposits its energy in a scintillator, tens of thousands of photons are produced, and tracking them across the detector takes a long CPU time. As a result, on machines with numerous cores or CPUs, running in multi-threading mode dramatically improves simulation speed.

NuSD can be run in two different modes, one in interactive mode with visualization and the other in batch mode without visualization. The former is typically used in the preliminary stages of the simulation to verify the detector set-up, the model, and the typical outcome of an event. On the other hand, the latter is much faster and is generally used to accumulate statistics. 

NuSD is written in such a way that it can be easily extended. The intent here is to easily incorporate a new detector concept into the program. Furthermore, the code is written in a simple language without using advanced topics of C++ that anyone with basic knowledge of C++ and OOP can readily understand and make changes on it.   

\subsection{Construction of NuSD detectors}
\label{sec2.2}

NuSD currently models seven different detector concepts based on the respective experiments. In addition to the G4VUserDetectorConstruction class provided by Geant4, NuSD introduces two new intermediate abstract detector construction classes, NuSDVDetConstruction and NuSDVCompositeDetConstruction, to derive a NuSD detector. The former is inherited from the G4VUserDetectorConstruction class and is used to create detector types that use a single scintillator. The latter is inherited from the first one and is used to construct composite detector types. 
 
\begin{figure}
 \centering
    \includegraphics[width=\linewidth]{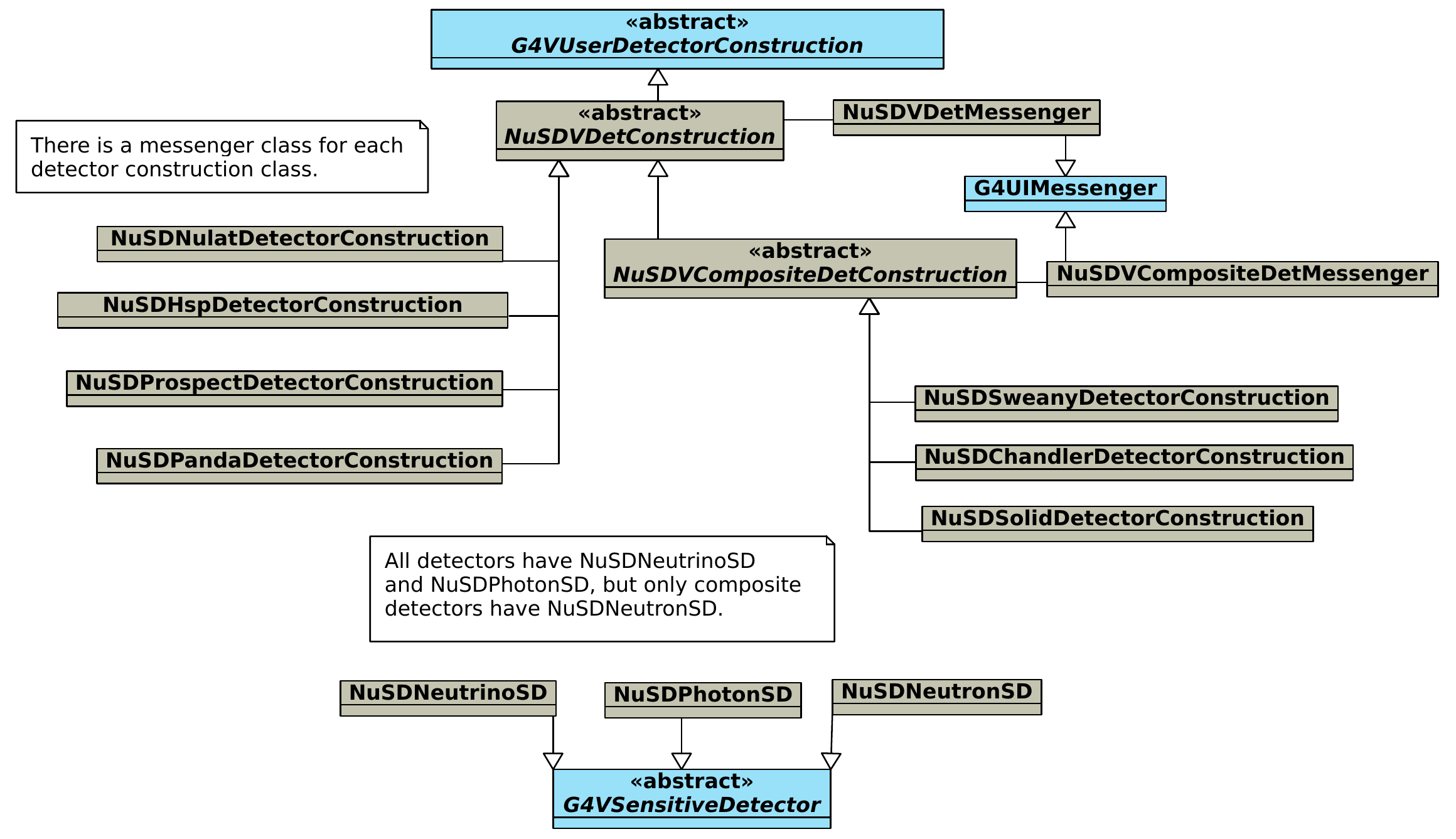}
     \caption{ The UML diagram of the classes involved in building NuSD detectors.  }
\label{fig:fig2}
\end{figure}

These two abstract classes contain common attributes and implement common functionality of all derived detector construction classes. Fig. \ref{fig:fig2} depicts the UML diagram of a detector construction process in NuSD.

A NuSD detector roughly consists of a main volume, which is in the form of a \textit{scintillation lattice}, and a number of identical optical readout units surrounding it. Each readout unit contains a photosensor, and optionally a light guide and accessory elements that act as glue between the scintillator and light guide or the scintillator and the photosensor.

To construct a detector in NuSD, first, the smallest repeating unit of the main volume of that detector is created. Then, in a loop, these unit volumes are positioned in a certain geometrical arrangement within a mother volume that represents the entire volume of that detector. In this way, the main part of the detector is formed. The optical readout system can also be installed together with the main volume within the same loop, according to the user's request.

NuSD detectors differ from each other in various aspects. First, each detector type has its own distinct repeating units, and the techniques for combining these units may differ. The differences can be grouped under five headings:

\begin{itemize}
\setlength\itemsep{0.em}
\item Optical readout system: PMT and SiPM.     
\item Neutron converter materials. Different reactions for neutron detection. Neutrons are converted into charged particles via Li-6 or gamma rays with Gadolinium (Gd).
\item The method of incorporating neutron converter materials into detectors. 
\item Detector segmentation: both physical and optical. 
\item Type of scintillators: plastic and liquid.
\item Signal identification techniques: pulse shape discrimination with appropriate material selection and topological selection cuts via segmentation.
\end{itemize}

\begin{figure}
 \centering
    \includegraphics[width=\linewidth]{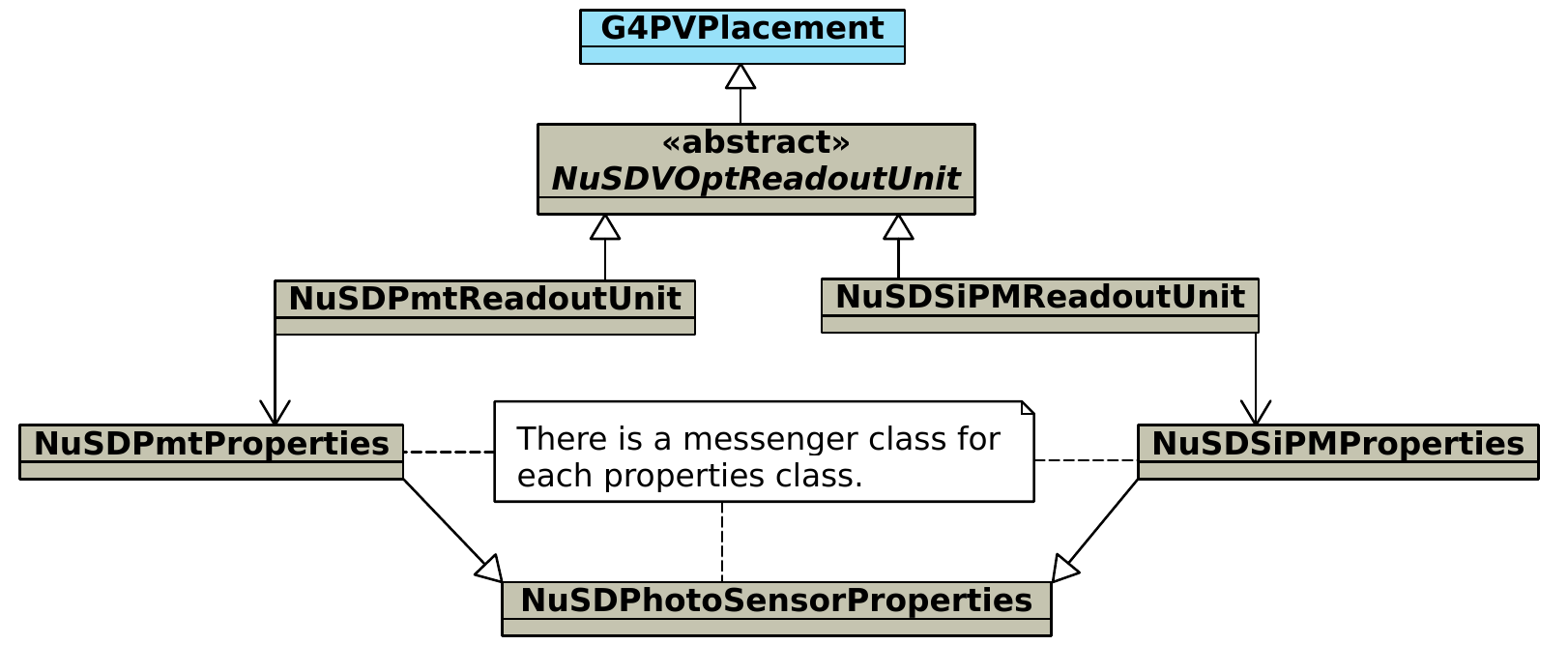}
     \caption{The UML diagram of the classes involved in building a major detector component. In this example figure, it is an optical readout unit.   }
\label{fig:fig3}
\end{figure}

NuSD uses two different ways to construct a detector component. If it is a tiny component, that is, only a few lines of code are required to produce it, then it is created in its own detector construction class. If it is a major component that necessitates a significant amount of code, a new class is derived from the G4PVPlacement, and all steps connected to this component are implemented in this new class. As an example, Fig. $\ref{fig:fig3}$ depicts the classes used to construct an optical readout unit, a key component common to all detectors, and their relationship in UML notation.

All NuSD detectors are constructed as described above. Therefore, if a new detector concept is to be incorporated into NuSD, using a similar approach is recommended in terms of compatibility.

Since each detector type may have its unique set of parameters, a messenger class is created for each detector construction class, inheriting from the G4UIMessenger. While the NuSDVDetMessenger class contains commands common to all detectors, the NuSDVCompositeDetMessenger class contains commands that are specific to composite detectors.  
All of the commands defined in a detector messenger class are kept in an ASCII file with the same name as that of the detector (i.e. Chandler.mac, Solid.mac...). These macro files allow a selected detector type to be set up in a desired geometric configuration. The following lists some critical parameters that can be adjusted when setting up a NuSD detector:

\begin{itemize}
\setlength\itemsep{0.em}
\item Number of segments along x, y and z.
\item Dimension of the neutrino scintillator.
\item Thickness of coating and container materials. For example, neutron converter sheet, optical barrier material, reflector, and aluminium tank.
\item Dimension of light guide and accesory elements (such as cement and grease). 
\item A flag to add/remove optical readout system.
\item A few true/false flags to include/remove some detector components.
\item A flag to indicate whether photosensors will be installed on either side of the detector.
\item Parameters related to the photosensors (PMT or SiPM) and reflector. 
\end{itemize}

In NuSD, the scintillator that detects neutrinos is referred to as a neutrino scintillator, whereas the scintillator that detects neutrons is referred to as a neutron scintillator. Neutron scintillators are only available in composite detectors, whereas neutrino scintillators are employed in all detector types.

The sensitive detector class in Geant4 has the task of creating hits each time a particle traverses a sensitive volume. To retrieve information from the simulation, the user writes a concrete sensitive detector class, inherited from G4VSensitiveDetector, that produces hits, and attaches it to the volume in the detector geometry from which the information is desired.

NuSD utilizes sensitive detector approach to retrieve information from the following three components:

\begin{itemize}
\setlength\itemsep{0.em}
\item NuSDNeutrinoSD: is attached to the neutrino scintillator volume of the detector. 
\item NuSDNeutronSD: is created only for composite detectors and is attached to the neutron scintillator portion of the detector. 
\item NuSDPhotonSD: is attached to the photon detection area of the optical readout system.
\end{itemize}

As an important point, NuSDPhotonSD cannot be triggered like a normal sensitive detector since it does not allow photons to pass through it. To trigger it, the status of the G4OpBoundary is monitored whenever a photon hits the photo-sensitive area of a photosensor (i.e., photo cathode or mppc). If the status is 'Detection', NuSDPhotonSD objest is retrieved from the G4SDManager and its ProcessHits function is called manually.

\subsection{NuSD materials}
\label{2.3}

Material definitions in Geant4 can be done in two alternative ways: the first one utilizes the G4NistManager class, which implements an interface to the National Institute of Standards and Technology (NIST) material database, allowing users to retrieve predefined isotopes, elements, and materials. Alternatively, the users can build their own custom materials from scratch using the provided G4Element, G4Isotope, and G4Material classes.

NuSD uses both approaches for material definition through the use of a singleton class named NuSDMaterials. It keeps a pointer to the G4NistManager and has a public method GetMaterial that takes a single parameter specifying the name of the material. This method internally invokes FindOrBuildMaterial method of G4NistManager and/or GetMaterial method of G4Material class. All materials utilized in the simulation are defined within the NuSDMaterials, and when a material is needed for a detector component, the GetMaterial method of the NuSDMaterials is called from the corresponding detector construction class. In this manner, materials common to distinct detectors are defined only once, rather than being defined repeatedly in each detector construction class.

The materials that make up the components of NuSD detectors, as well as their functions in a detector, are outlined below:

\begin{itemize}

\setlength\itemsep{0.em}
\item Several commercially available scintillators from Eljen Technology\textsuperscript{\textregistered} such as EJ-200, EJ-260, EJ-335, and EJ-390 and two custom-made scintillators developed for Prospect and Nulat experiments, Custom-EJ-309 and Custom-EJ-254. These are referred to as neutrino scintillators in NuSD, and a neutrino scintillator is the primary component of a neutrino detector. 

\item Thermal neutron scintillator screen. It is composed of a matrix of ZnS:Ag and Li-6. In addition to a neutrino scintillator, composite detectors employ a neutron scintillator. It converts neutrons into charged particles. 

\item Gadolinium coated Mylar films. It converts neutrons into gamma rays.
	
\item Optical barrier materials such as air, FEP (Fluorinated ethylene propylene), and water. The main function of these materials is to direct the photons emitted from the scintillator towards the light collectors through total internal reflections. Therefore, materials with low refractive index are prefered.

\item Reflector sheets such as 3M reflector film and aluminized mylar film. It is used to improve light collection and also for optical segmentation.
\item PMMA (polymethylmethacrylate). It is also known as acrylic, acrylic glass, perspex, or plexiglass. Light guides are fabricated from acrylic materials. A light guide is generally used to improve energy resolution of a detector.
\item Accessory elements such as cement and grease. The former is used to bind a scintillator to a light guide, while the latter couples a scintillator directly to a photosensor.	
\item Containter materials such as stainless steel tank or acrylic.
	 
\end{itemize}

These are the common materials used to construct a complete neutrino detector. The users can easily define new materials in the same way as in NuSDMaterials class.

A material is defined by specifying its density, the number of components, and the fraction of each component that makes up the material. The specification of the isotope composition of the material is especially significant for the hadronic interactions since the hadronic cross-section is a strong function of the nuclear property of the material. If the physics processes of a simulation do not involve optical interactions (or the optical processes are disabled by the user), no additional material information is required. If it involves, optical properties of medium which are key to the implementation of the optical processes, should be defined.

To accurately simulate optical photons, the user must define all material and surface properties. Many parameters are provided by Geant4 to accomplish this. The definitions of these parameters and how these parameters are implemented are explained in detail in the Application Developer's Guide \cite{G4App}.

To put it briefly, Geant4 provides the G4MaterialProperties table class to impart new properties to a pre-defined material or a surface. For this, an instance of G4MaterialPropertiesTable is created, and the desired number of properties is added to the table as entries. This table showing properties is then linked to the material or surface in question. Each property in the table may be independent of energy (denoted as "Constants") or it may be expressed as function of the photon energy. If a property does not depend on energy, it has a single value. If it depends on energy, it contains a list of energy-value pairs.

NuSD utilizes a set of text files to store the energy-value pairs of energy-dependent quantities. The text files can be arranged in either one or two columns. The single-column text files include two discrete energy values that indicate the energy response range of the interested quantitiy. In this case, the same default value is used at all energies. Using the same value at all photon energies for a time-dependent quantity enables to directly measure the impact of that property on the interested quantity. On the other hand, the two-column text files contain a list of energy-value pairs. A NuSD user first decides whether a property depends on energy or not, and then edits the relevant text file accordingly. In this way, a NuSD user can run the simulation program using his/her own experimental input parameters.

The provided text files are kept in three separate folders that are grouped according to their contents:

\begin{itemize}
\setlength\itemsep{0.em}
\item Quantum efficiency spectra of the used photo sensors (photocathode for PMT, and silicon microcells for SiPM/MCPP). NuSD uses NuSDPhotoSensorSurface class for the implementation of a photo sensor surface. 
\item Reflectivity spectra of the used reflectors. NuSD uses NuSDReflectorSurface class to implement a reflector surface.
\item Refractive indices and absorption lengths of the materials, as well as the photon emission spectra of the used scintillators. 
\end{itemize}

These three folders are stored under a parent directory called \textit{data}, which is in the same path as the source code of NuSD. It is the user's responsibility to search and find these experimental input values that are very crucial for accurate simulation. For example, Eljen technology provides data sheets for the scintillators it produces. A user can extract the relevant information from the supplied data sheets by the manufacturers. 

\subsection{Physics processes}
\label{sec2.4}

Physics processes describe how particles interact with materials. Geant4 uses the concept of physics lists to handle the physics of the simulation. A physics list is a collection of physics processes, models, and cross-sections that are used to model the process of particle passage through matter.

There are three ways to build a physics list in Geant4. In the first case, the user has to manually specify all the particles to be used in the simulation as well as the physics processes assigned to each particle. Although it provides a flexible way to set up the physics environment, it requires deep knowledge of the physics of the target application. The second one is the usage of the "physics constructor" which allows to group particles and their processes according to a physics domain. A physics constructor handles a well-defined category of physics (e.g EM physics, hadronic physics, decay, etc). The third option is the "reference physics lists" which cover all the physics interactions that take place in the simulation. These are ready-to-use complete physics lists provided by the toolkit and are routinely validated and updated with each release. Geant4 provides several reference physics lists that are suited for different use cases.

NuSD uses QGSP-BERT-HP/QGSP-BIC-HP (“HP” for high-precision neutron interactions) reference physics list for the simulation of IBD events. The QGSP-BERT-HP includes standard electromagnetic and hadronic physics processes and is specifically suited for the transport of neutrons below 20 MeV down to thermal energies \cite{QGSP}. This modular physics list covers all the well-known processes such as ionization, Coulomb scattering, Bremsstrahlung, Compton scattering, photoelectric effect, pair production, annihilation, decay, radiative capture, fission, hadronic elastic, and inelastic scattering. This reference physics list does not include optical processes. To include them, G4OpticalPhysics, the constructor of optical processes, is instantiated and then registered in the list. This automatically adds all the optical physics processes.

\subsection{Primary particles}
\label{sec2.5}

\begin{figure}
    \centering
          \includegraphics[width=\linewidth]{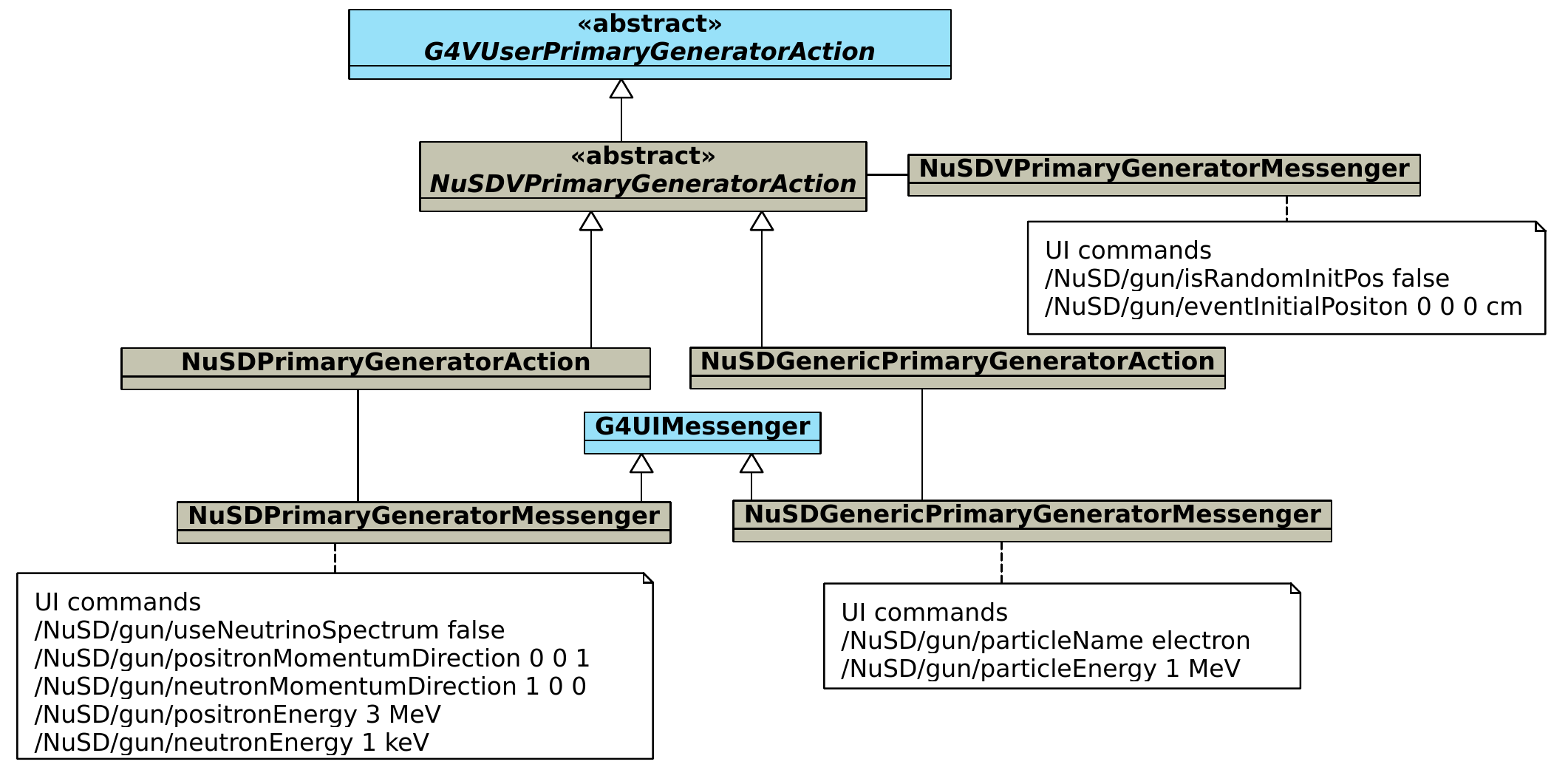}
          \caption{The UML diagrams of the classes involved in generating primary vertex and primary particles. If the command \textit{/NuSD/gun/useNeutrinoSpectrum} is set to true, the commands below it are ignored.}
    \label{fig:fig4}
\end{figure}

NuSD provides two different primary particle generation classes: NuSDPrimaryGeneratorAction and NuSDGenericPrimaryGeneratorAction. The first one is used to generate IBD events, while the second one is used to simulate background events or any desired particle. These two classes are inherited from the NuSDVPrimaryGeneratorAction abstract class and are used to identify the primary particles to be generated as well as setting their initial values for energy, momentum, position, and time. The NuSDVPrimaryGeneratorAction class itself is derived from the G4VUserPrimaryGeneratorAction class and includes some additional member functions for generating random positions within the active volume of NuSD detectors. Fig. $\ref{fig:fig4}$ shows the UML diagrams of the primary generator classes used by NuSD.

The initial position of an event can be specified in one of two ways in NuSD. The first one is to generate an event at a random point inside the detector, and the second one is to generate an event at a user-specified position. In the first option, which is accomplished by setting the command /NuSD/gun/isRandomInitPos to true, a random unit is first selected, then a random point is generated within that unit. If the command is set to false, a random unit is selected again, but this time the position is specified relative to the selected unit's center via the command /NuSD/gun/eventInitialPositon. 

NuSD generates an IBD event by directly creating its products as primary particles due to the neutrino's extremely low interaction cross-section. By default, NuSD simulates reactor neutrinos (1-10 MeV energy range) and performs the following steps to determine the initial energy of the positron and neutron:

\begin{itemize}
\setlength\itemsep{0.em}
  
\item Eq. $\ref{eq:0}$ is used to sample neutrino energies. Here $P(E_{\overline{\nu}})$ is the detecting probability of a reactor neutrino at energy $E_{\overline{\nu}}$ , $\Phi_i (E_{\overline{\nu}})$ is the emitted reactor neutrino energy spectrum per fission of each of the four isotopes, $\sigma (E_{\overline{\nu}})$ is the neutrino-proton interaction cross section at zeroth-order,  $\alpha_i$ is the average fission fraction of each isotope over the reactor fuel cycle, and N is the normalization constant.

\item Eq. $\ref{eq:1}$ is used to sample the cosine of the scatterring positron angle \cite{Vogel}. 

\item Using the ROOT analysis framework, two histograms, one for the initial neutrino energy and the other for the scattering positron angle, are generated and written into a ROOT file.  

\item The NuSDPrimaryGeneratorAction class uses this ROOT file to get a neutrino energy and a positron scattering angle from the written histograms for each event.

\item The positron energy $E_{e^+}$ is computed from Eq. $\ref{eq:2}$. Here $m_e$ is the electron mass, M is the neutron mass, and the $\triangle$ is the mass difference of the neutron and proton.     

\item The neutron energy $T_n$ is calculated from Eq. $\ref{eq:3}$. 

\end{itemize}

\begin{eqnarray}
\label{eq:0}
 P(E_{\overline{\nu}}) &=& \frac{1}{N}  \sum_{i=5,8,9,1} \alpha_i  \Phi_i (E_{\overline{\nu}}) \sigma (E_{\overline{\nu}}) \\
\label{eq:1}
P(\cos\theta)  &=& \frac{1}{N} \left[ 1- 0.1 \cos \theta \right]   \\
\label{eq:2}
E_{e^+}^{(1)} &=& E_{e^+}^{(0)} \left[ 1- \frac{E_\nu}{M}\left(1- \cos\theta \right) \right]-\frac{\triangle ^2-m_e^2}{2M} \\
\label{eq:3}
T_n &=& \frac{ E_\nu \left(E_\nu -\triangle\right) }{M} \left(1-\cos\theta \right) + \frac{\triangle ^2-m_e^2}{2M}
\end{eqnarray}

As an alternative to the above, a NuSD user can manually set the initial values of the positron and neutron with the supplied user interface commands (see Fig. $\ref{fig:fig4}$).

The NuSDGenericPrimaryGeneratorAction class is implemented to simulate electrons by default. The user can alter the type of particle and its initial setting using the supplied user commands.

\begin{figure}
\centering
 \includegraphics[width=\linewidth]{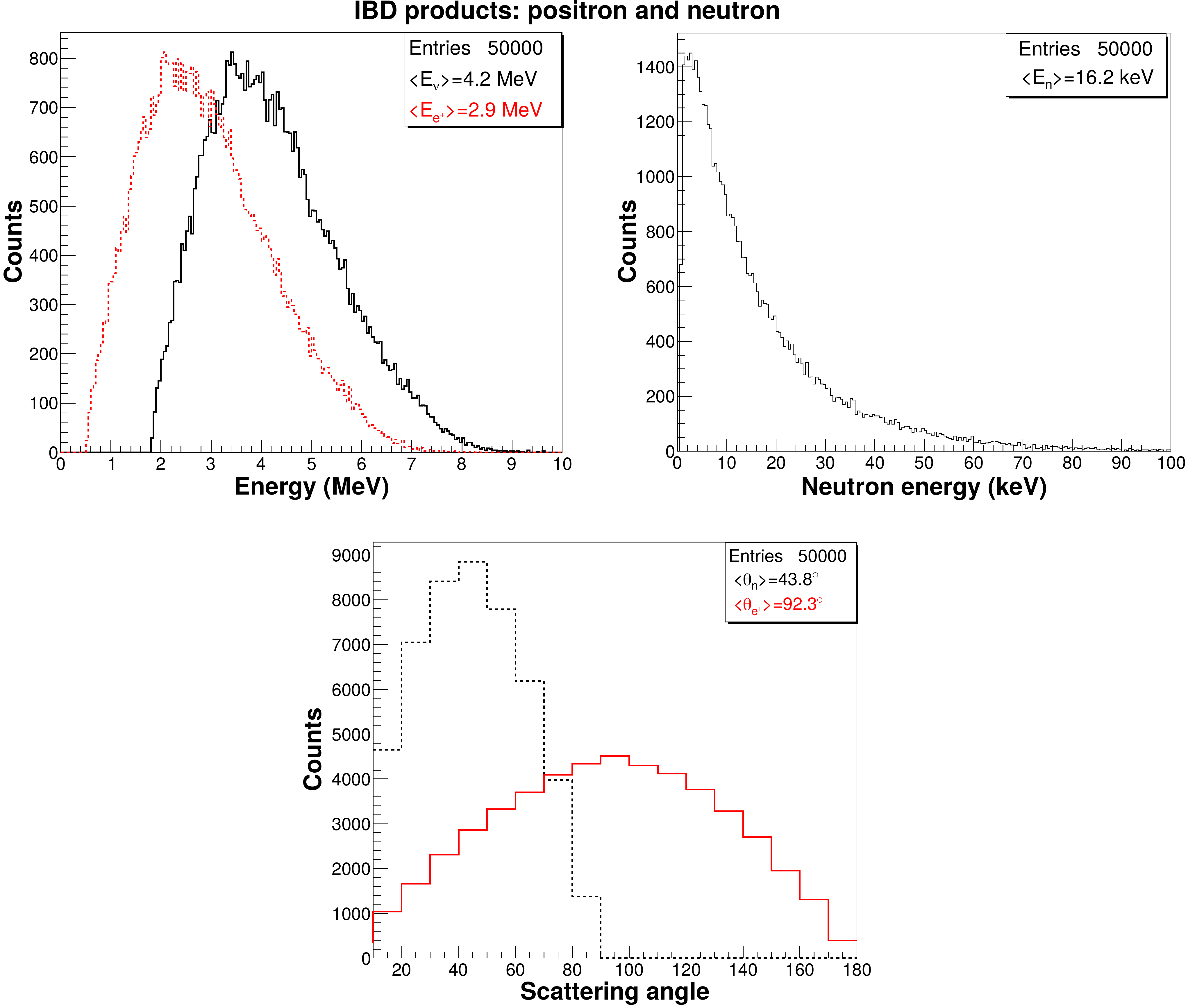}
 \caption{ The initial energy distributions of the positron (top-left) and neutron (top-right). The scattering angle distribution of the positron and neutron (bottom). }
 \label{fig:fig5}
\end{figure}

Fig. $\ref{fig:fig5}$ shows the initial energy distributions of the positron and neutron at the top, and their scattering angle distribution at the bottom. From the bottom of the Figure, it can be seen that the average positron direction is slightly backward and the neutron direction is purely forward.

\subsection{Output of NuSD}
\label{sec2.6}

To extract information from the simulation, Geant4 provides the g4tools package and analysis manager classes that can be used to write data in several formats (ROOT, AIDA XML, CSV and HBOOK). NuSD prefers ROOT output technology since it is well-documented and is the most widely used tool for data analysis in the High Energy Physics community.

NuSD collects information broadly in two ways. Firstly, through the use of sensitive detector classes (NuSDNeutrinoSD, NuSDNeutronSD, and NuSDPhotonSD) that generates hits, which contains some desired information. The information acquired from the sensitive detectors is common for both IBD and single-particle simulation. Secondly, via user action classes. User action classes give users complete control over the simulation and access information from various stages of the simulation.

Both methods mentioned above are handled by use of two singleton analysis manager classes, NuSDAnalysisManager and NuSDGenericAnalysisManager. The former is used for the outputs of the IBD events, while the latter is used for any other events to be simulated. The users of NuSD can rearrange NuSDGenericAnalysisManager class depending on the desired outputs. 

The following physics quantities are obtained from the IBD on an event-by-event basis, and each is recorded in a different column of an Ntuple.

\begin{itemize}
\setlength\itemsep{0.em}

\item Event ID
\item Antineutrino energy that initiates the IBD event, the initial energy of the IBD products (i.e positron and neutron), and the momentum direction of these three.
\item Event vertex position (3D vector in Cartesian coordinates).
\item Neutron capture position (3D vector in Cartesian coordinates).
\item Neutron capture time. 
\item Mass number of the neutron-capturing nucleus.  
\item Number of emitted scintillation photons.
\item Sensitive detectors create hits from step and store them into hits collection. A hit object created by the NuSDNeutrinoSD or NuSDNeutronSD has the following attributes: particle PDG code, track ID, copy number, deposited energy, and energy deposition time. At the end of the event, a std::vector is created for each member of the hit.  
\item The hit object created by the NuSDPhotonSD for each detected photon has two members: the detection time of the photon and the copy number of the PMT/SiPM that detects that photon. At the end of the event, a std::vector is created for both members of the hit.  
\end{itemize}  

Although NuSD is not an analysis framework, it provides several ready-to-use ROOT macro files to analyze raw data obtained from the simulations. These macro files plot the simulation data and calculate some important quantities that cannot be obtained directly from the simulations. Some of those are:

\begin{itemize}
\setlength\itemsep{0.em}

\item Total deposited energy in each segment of the detector. In other words, energy deposition map of a detector.
\item Number of detected photons by each photosensor in an event.
\item Neutron capture efficiency of nuclei such as hydrogen, carbon, Li-6, and Gd isotopes. 
\item Mean neutron capture time.
\item Antineutrino detection efficiency. 
\item Light collection efficiency.

\end{itemize} 

The provided macro files enable users to apply several selection cuts when calculating a quantity. For example, cuts such as the amount of deposited energy and its deposition time, and detected photon number and its detected time.

\section{NuSD Workflow}
\label{sec3}

The following summarizes the main steps a user follows when running a NuSD application:

\begin{itemize}
\setlength\itemsep{0.em}

\item When the program is built, a file entitled \textit{NuSD\textunderscore config.h} is generated in the build directory. This file contains a few settings that are changed rarely. Using this file, configure the initial simulation settings that are explained in Sec. \ref{sec2.1}. 

\item There is a macro file for each detector type under the folder "macros" in the build directory. Using the relevant macro file, set up the selected detector with desired geometric layout. 

\item Next, start NuSD in interactive mode to visualize the installed detector and then monitor the event to be simulated by pressing the run button. This mode is typically used at the initial stage of the simulation for the validity of the detector set-up and to see the behavior of particle tracks at the surfaces. For example, one can reveal the number of photons hit on a photosensor or the number of photons absorbed by an optical surface by watching photons on those surfaces with naked eyes. Further, an event summary will be be printed on the terminal at the end of the event if NuSD\textunderscore DEBUG macro is enabled. The geometry can also be altered at this stage (Idle state) using the supplied interface available on the panel.

\item Once the validation stage is successfully completed, run the program with the desired number of events in batch mode to accumulate statistics. Working in multi-threading mode is recommended if optical photons are simulated.

\item When the program has finished running, ROOT files will be created automatically if the CREATE\textunderscore ROOT\textunderscore FILE macro is enabled.

\item Using the supplied ROOT files, make your analysis. Many analysis ROOT macros have already been provided for users.

\end{itemize}

\section{Results}
\label{sec4}

This section is divided into three parts. The first part briefly introduces and categorizes the detectors modeled using NuSD. The second part presents some simulation results over a selected detector type. The final part identifies a few key design evaluation metrics that are critical for all detector types, and findings are reported using these quantitative metrics.

\subsection{NuSD detectors}
\label{sec4.1}
\begin{figure*}
 \centering
    \includegraphics[width=0.9\textwidth]{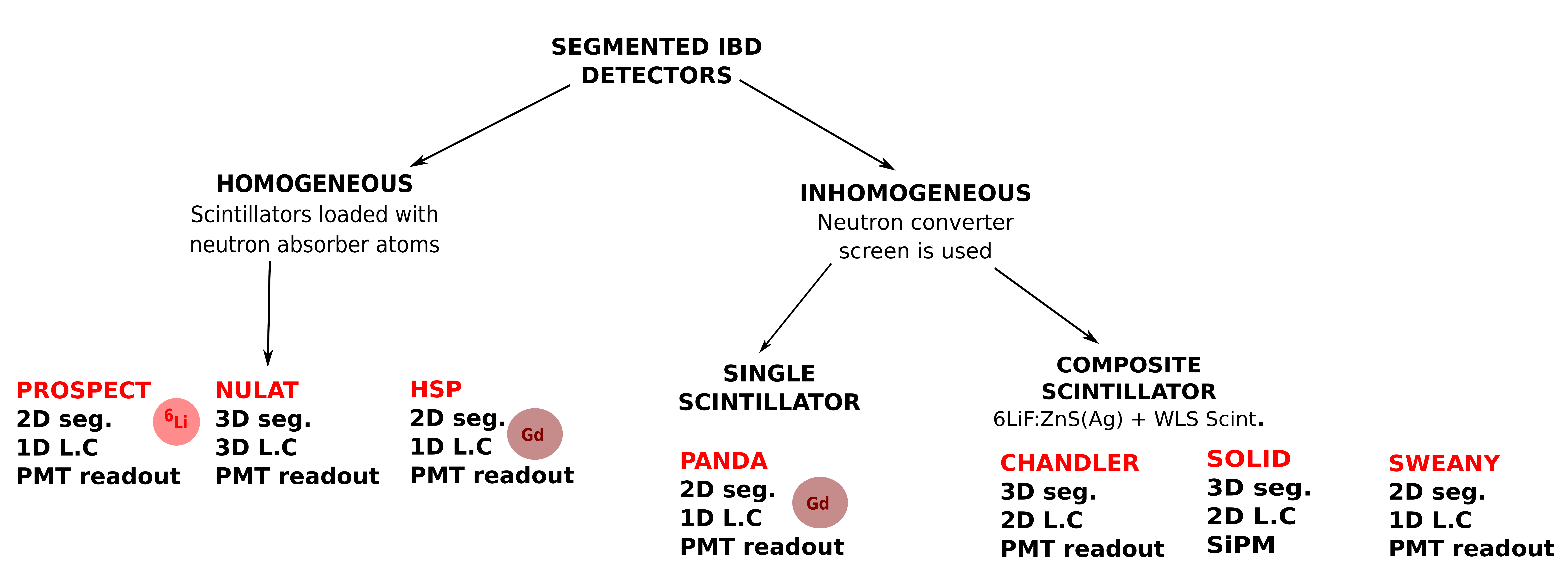}
     \caption{Detector classification based on the selection of scintillation material and the application method for neutron detection. }
    \label{fig:fig6}
\end{figure*}

We classify segmented IBD detectors as homogeneous and inhomogeneous. In a homogeneous detector, the neutrino scintillator is uniformly loaded with a neutron capture agent, while in an inhomogeneous detector, the neutron capture material is in the form of a thin foil/sheet and is wrapped around the scintillator. Moreover, depending on whether a single type or dual type scintillator is employed, inhomogeneous detectors are split into two types. Fig. $\ref{fig:fig6}$ shows the categorized NuSD detectors schematically.

\subsubsection{Homogeneously doped detectors}
\label{sec4.1.1}
\paragraph{NULAT-Style}

NuLat (short for Neutrino Lattice) is a segmented detector design based on the Raghavan Optical Lattice (ROL) concept. Its first phase consists of a cubical assembly of 125 $^6$Li-doped plastic scintillator cubes (5$\times$5$\times$5 cm$^3$) and the entire detector will be made of 3375 cubes (15$\times$15$\times$15). The air gaps separating each cube enhance the total internal reflection effect thus channeling the light emitted in each cube after an energy deposition towards photosensors placed on the sides of the detectors. This concept provides an excellent spatial and energy resolution within the entirety of the detector and is well-suited to reject external backgrounds.

\begin{figure}
    \centering
        \includegraphics[width=\linewidth]{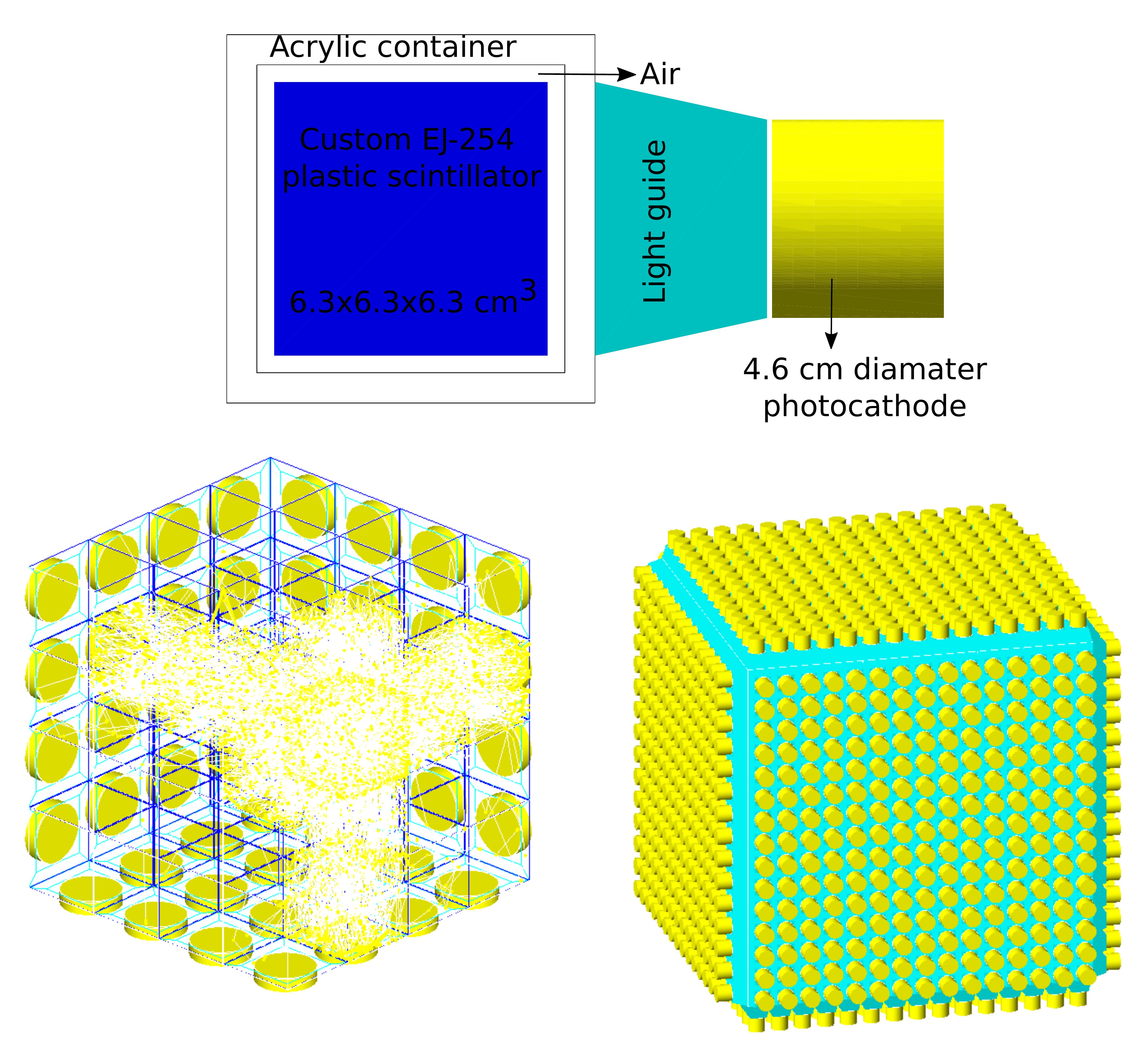}
    \caption{NULAT-like detector concepts \citep{Nulat:2015,Nulat:2018}. 3D segmentation and 3D light collection that is based on the Raghavan Optical Lattice (ROL) \cite{Lens:2007}. }
    \label{fig:fig7}
\end{figure}

Fig. \ref{fig:fig7} shows three different layouts of a Nulat-style detector plotted with NuSD: a single module with a dimension of 6.3$\times$6.3$\times$6.3 cm$^3$ (top), an example configuration of 4$\times$4$\times$4 showing how light is transported (bottom-left), and finally a version of 15$\times$15$\times$15 corresponding to 0.85 m$^3$ active volume (bottom-right).

\paragraph{PROSPECT-Style}

The Precision Reactor Oscillation and Spectrum Experiment (PROSPECT) detector design consists of optically isolated segments of $^6$Li-doped liquid scintillators. The entire lattice is made of 11$\times$14 segments, 1.2~meters in length, for a total weight of about 4~tons. Due to the reflective internal walls of the segments, light emitted after an energy deposition in the scintillator is propagated towards both sides of a segment to be detected by PMTs.

\begin{figure}
    \centering
    
        \includegraphics[width=\linewidth]{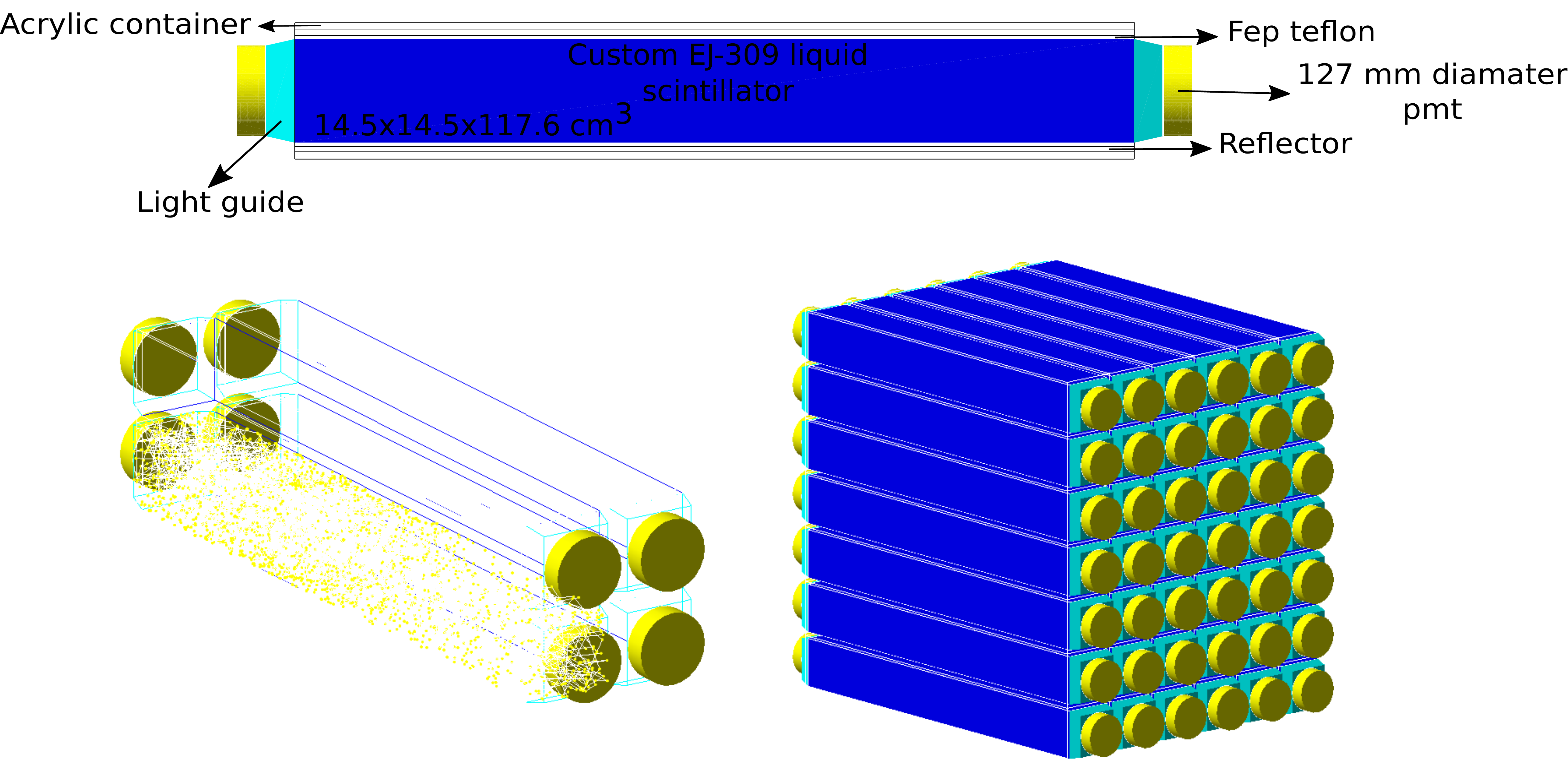}
        
    \caption{PROSPECT-like detector concepts \cite{Prospect:2015,Prospect:2018,Prospect:2019,Prospect:2019:2}. 2D segmentation and 1D light collection.}
    \label{fig:fig8}
\end{figure}

Fig. \ref{fig:fig8} depicts three different layouts of a Prospect-style detector plotted with NuSD: a single module with a dimension of 14.5$\times$14.5$\times$117.6 cm$^3$ (top), an example configuration of 2$\times$2$\times$1 showing how light is transported (bottom-left), and finally a version of 6$\times$7$\times$1 corresponding to 1.04 m$^3$ active volume (bottom-right).

\paragraph{HSP-Style}

\begin{figure}[htbp]
    \centering
        \includegraphics[width=\linewidth]{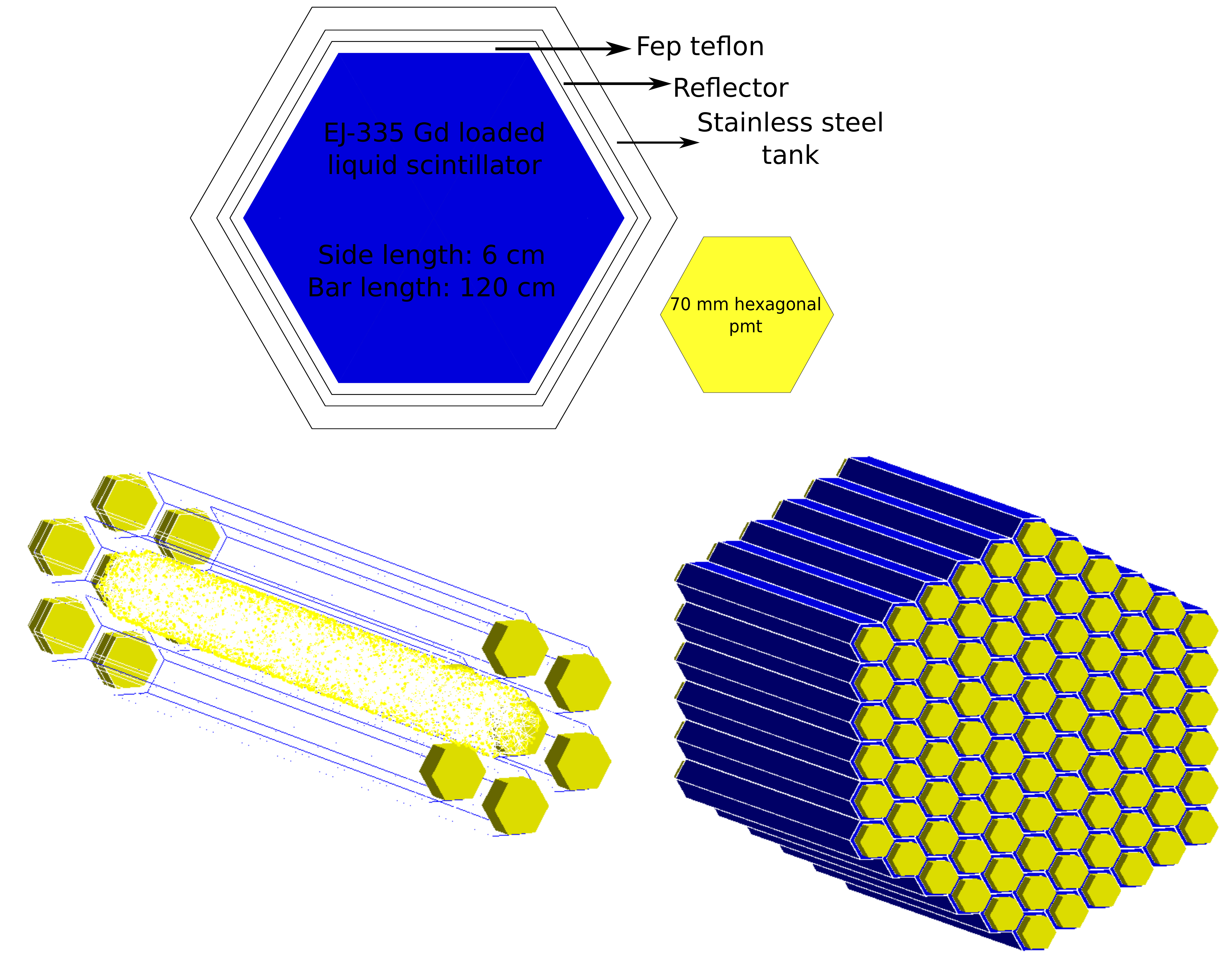}
    \caption{HSP detector concepts \cite{Hsp:2020}. 2D segmentation and 1D light collection.  }
    \label{fig:fig9}
\end{figure}

The Hexagonal Shaped Package (HSP) detector design is an assembly of 91 identical units in an hexagonal shape, as displayed in Fig.~\ref{fig:fig9}. Each unit is composed of an hexagonal volume, filled with liquid scintillator, with a side length of 6~cm and a height of 120~cm coupled on both sides to light guides and PMTs. The light collection efficiency is increased by wrapping the scintillator bars and the light guides with aluminized Mylar sheet. This final assembly offers an active volume of 1~m$^3$ and a gadolinium concentration of 0.25\% by weight.

\subsubsection{Inhomogenous detectors}
\label{sec4.1.2}

\paragraph{PANDA-Style}

\begin{figure}
    \centering
        \includegraphics[width=\linewidth]{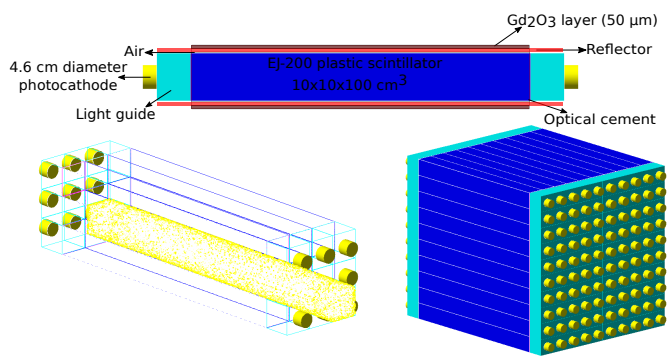}
    \caption{PANDA-like detector concepts \cite{Panda:2012,Panda:2014}. 2D segmentation and 1D light collection.}
    \label{fig:fig10}
\end{figure}

The Plastic Anti-Neutrino Detector Array (PANDA) concept consists of an array of 100 identical modules in a 10$\times$10 rectangular shape, as displayed in Fig.~\ref{fig:fig10}. Each module consists of a scintillator bar (10$\times$10$\times$100~cm) wrapped in aluminized Mylar film to increase light collection efficiency. A second layer of Mylar film is coated with a 50 $\mu$m-thick deposit of gadolinium oxide (Gd$_2$O$_3$)to enhance the neutron capture efficiency (4.9~mg of Gd per cm$^2$). Each bar is connected to acrylic light guides and PMTs on both sides.

\subsubsection{Inhomogenous composite detectors }
\label{sec4.1.3}

\paragraph{SWEANY-Style}

The SWEANY-like design, named after M.~Sweany, the corresponding author of the initial concept~\cite{Sweany:2015}, consists of a rectangular assembly of individual plastic scintillator units. Each unit is a 12.7$\times$12.7$\times$60 cm$^3$ bar covered on all long sides with a 0.45~mm thick layer of $^6$LiF:ZnS(Ag) scintillator and a PMT is coupled to each short side of the unit.

\begin{figure}
    \centering
    \includegraphics[width=\linewidth]{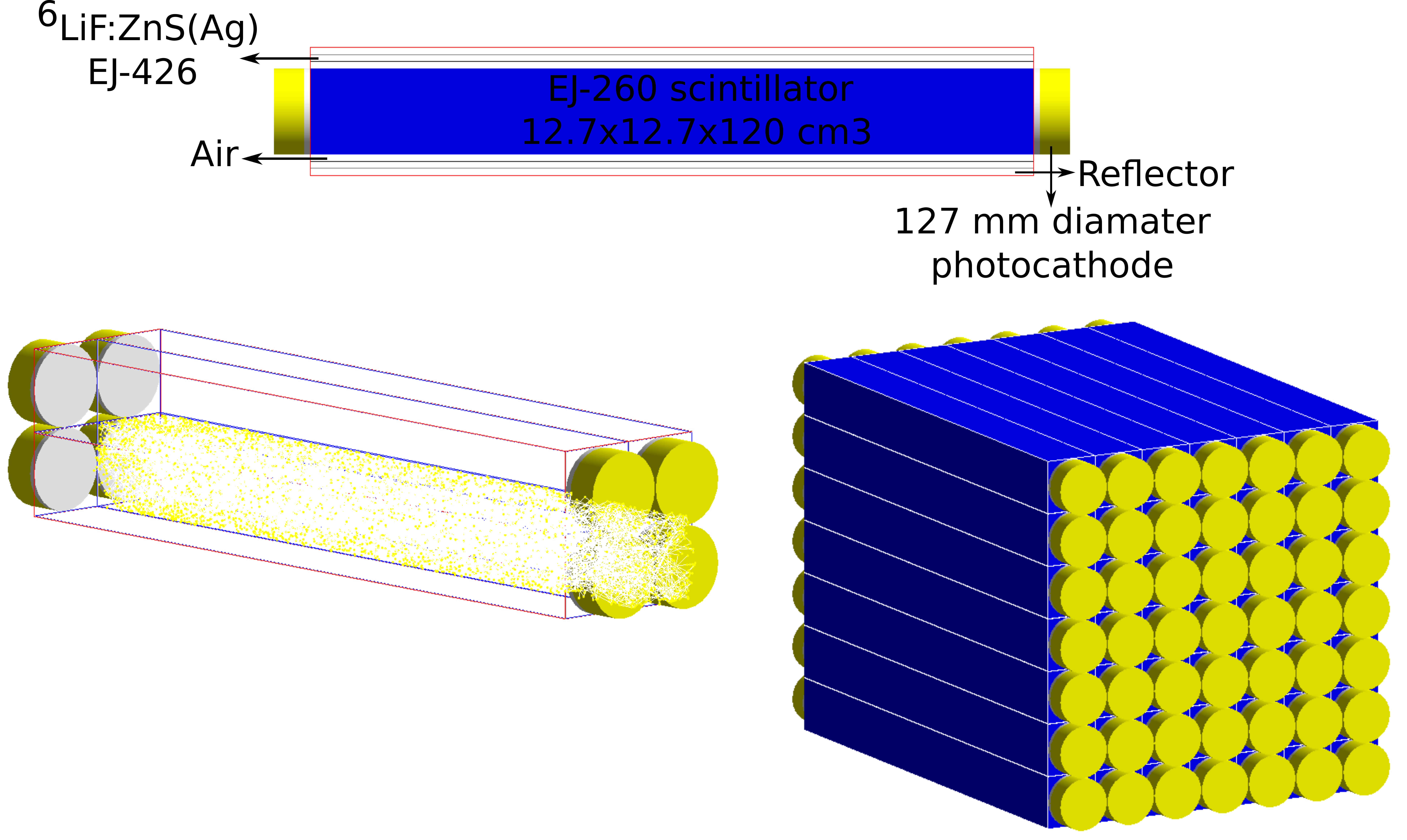}
     \caption{SWEANY-like detector concepts \cite{Sweany:2015,Scott:2011}. 2D segmentation and 1D light collection. }
    \label{fig:fig11}
\end{figure}

Fig. \ref{fig:fig11} shows three different layouts of a Sweany-style detector plotted with NuSD: a single module with a dimension of 12.7$\times$12.7$\times$120 cm$^3$ (top), an example configuration of 2$\times$2$\times$1 showing how light is transported (bottom-left), and finally a version of 7$\times$7$\times$1 corresponding to 0.95 m$^3$ active volume (bottom-right).

\paragraph{CHANDLER-Style}

\begin{figure}
    \centering
    
        \includegraphics[width=\linewidth]{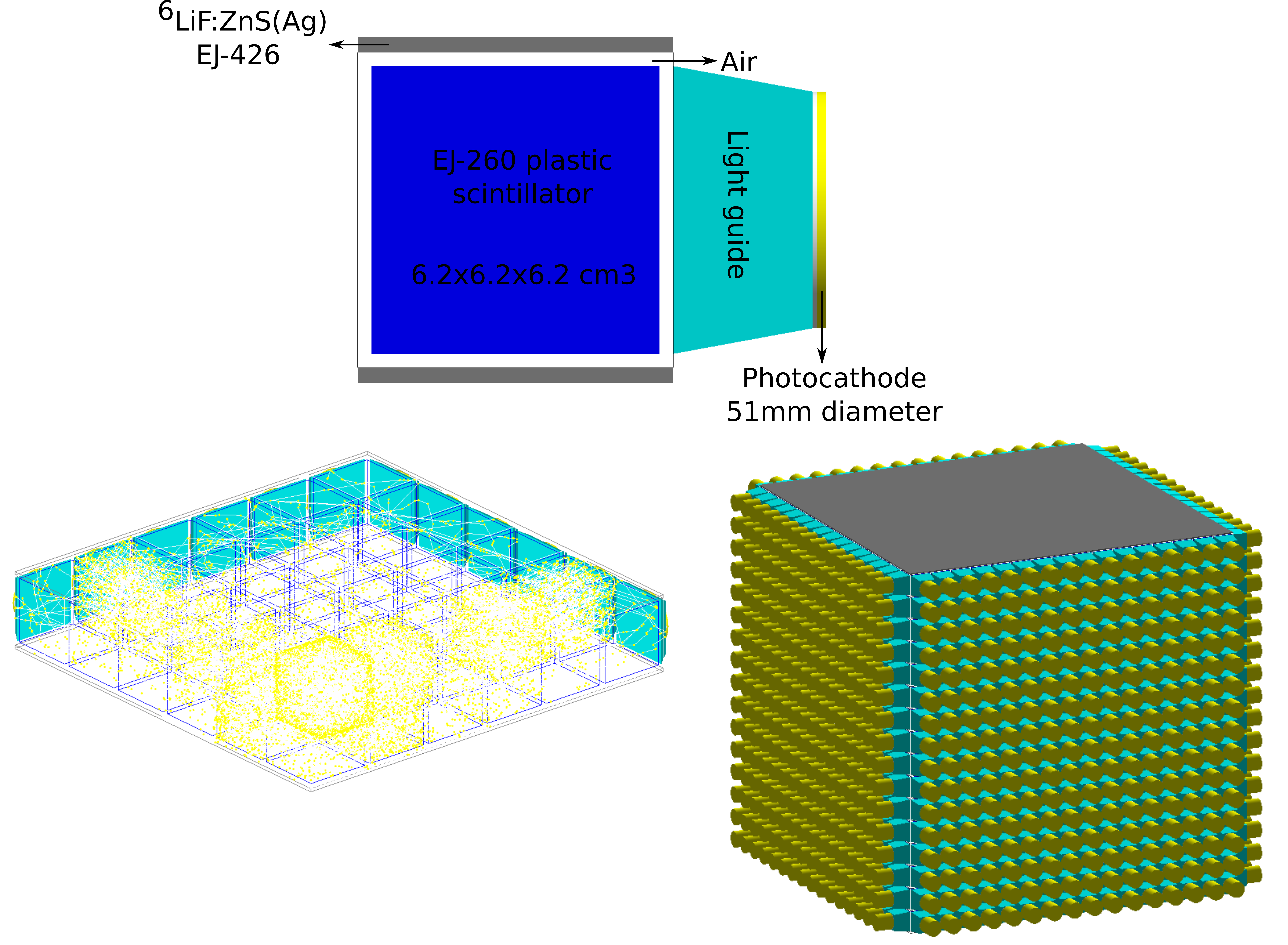}
        
    \caption{CHANDLER-like detector concepts \cite{Chandler:2018, Chandler:2019, Chandler:2020, Chandler2:2020}. 3D segmentation and 2D light collection that is based on Raghavan Optical Lattice \cite{Lens:2007}. Light is transported along rows and columns of cells by total internal reflection.}
    \label{fig:fig12}
\end{figure}

Similar to NuLat, Carbon Hydrogen Anti-Neutrino Detector with Lithium Enhanced Raghavan optical lattice (CHANDLER) is a segmented detector design based on the Raghavan Optical Lattice concept. The Raghavan Optical Lattice divides the active region of the detector into identical cubic cells. Each cube is made of wavelength shifting plastic scintillators. Light is transported along rows and columns of cells by total internal reflection. PMTs on both ends of the bars collect light generated in neutrino scintillators and neutron scintillators ($^6$LiF:ZnS). The neutron detection sheet optically isolates the cube layers, although a small amount of light can pass through the sheet.  
Fig. \ref{fig:fig12} shows three different layouts of a Chandler-style detector plotted with NuSD: a single module with a dimension of 6.2$\times$6.2$\times$6.2 cm$^3$ (top), an example configuration of 6$\times$1$\times$6 showing how light is transported (bottom-left), and finally a version of 16$\times$16$\times$16 corresponding to 0.98 m$^3$ active volume (bottom-right).

\paragraph{SOLID-Style}

Search for Oscillations with a $^6$Li detector (Solid) is a segmented composite scintillator technology which uses polyvinyl toluene (PVT) plastic scintillator cubes, each coupled to a neutron sensitive inorganic scintillator $^6$LiF:ZnS(Ag) screen. Each cube is also optically isolated from its neighbors, and the scintillation signals from the two scintillators are collected via the same wavelength shifting fibres connected with silicon Multi-Pixel Photon Counter. 

\begin{figure}
    \centering
    
        \includegraphics[width=\linewidth]{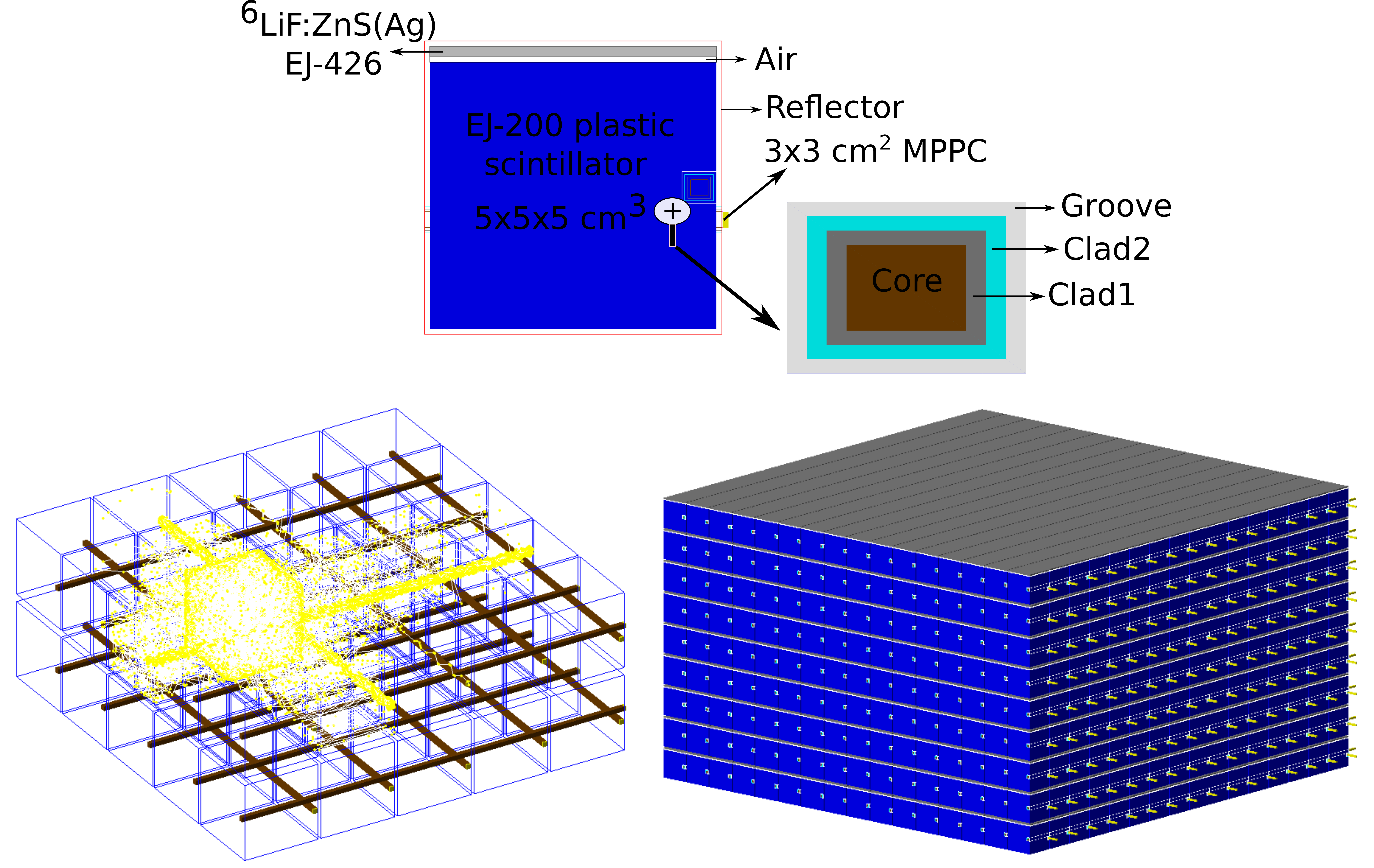}
        
 \caption{SOLID-like detector concepts \cite{Solid:2015,Solid:2017,Solid2:2017,Solid3:2017,Solid:2018,Solid2:2018}. 3D segmentation and 2D light collection.}
    \label{fig:fig13}
\end{figure}

Fig. \ref{fig:fig13} shows three different layouts of a Solid-style detector plotted with NuSD: a single module with a dimension of 5$\times$5$\times$5 cm$^3$ (top), an example configuration of 5$\times$2$\times$5 showing how light is transported (bottom-left), and finally a version of 16$\times$9$\times$16 corresponding to 0.29 m$^3$ active volume (bottom-right).

\subsection{Example}
\label{sec5.1}

We plot two important measurable quantities that are essential to the analysis of IBD data and cannot be obtained directly from the output of NuSD. The first one is an anti-neutrino signal that reveals itself with two correlated signals, the so-called prompt and delayed signal. The second one is a measurement of the total energy accumulated over all detector segments. Chandler is chosen as a representative detector and established based on the respective experiment.

\begin{figure*}
\centering
 \includegraphics[width=\linewidth]{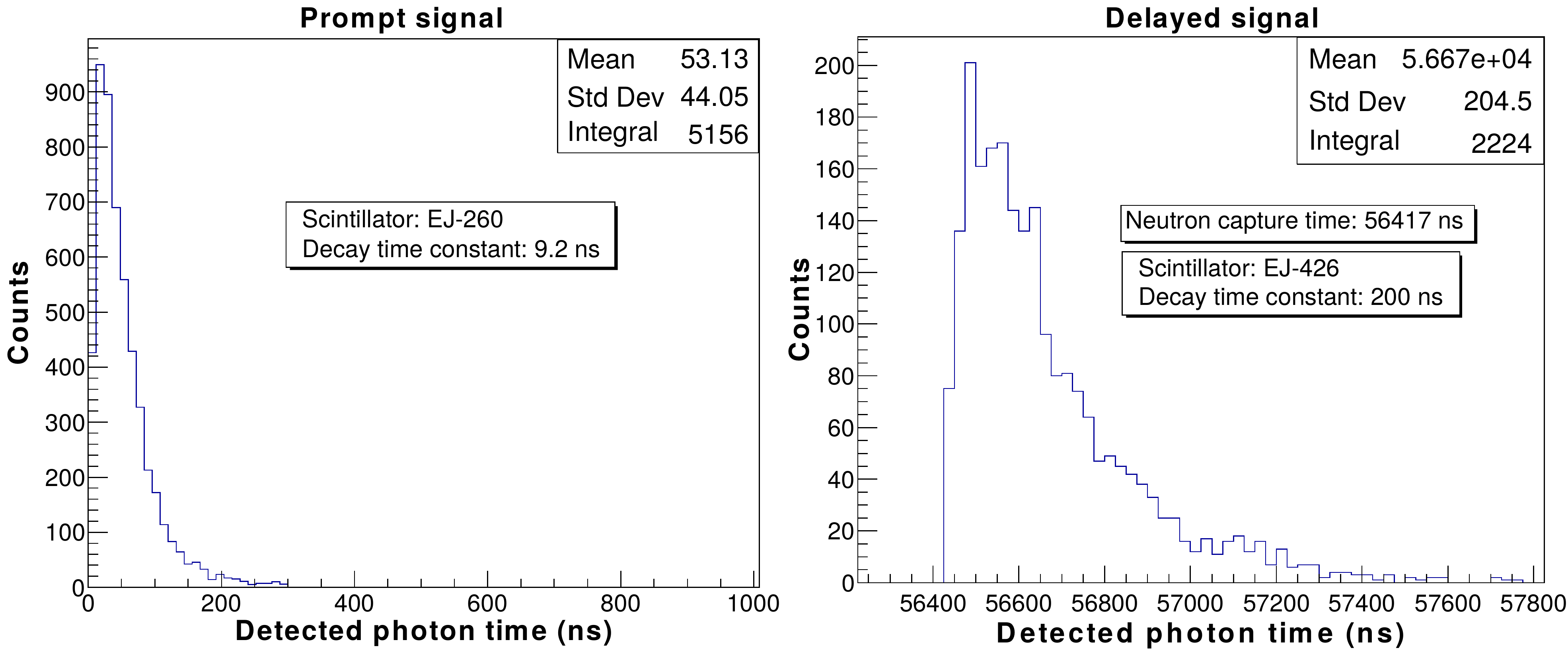}
 \caption{Prompt and delayed components of an anti-neutrino signal obtained with Chandler-style detector. Pulse shape discrimination is an effective method for discriminating prompt and delayed signals in this type of detector. }
\label{fig:fig14}
\end{figure*}

Fig. \ref{fig:fig14} depicts an exemplary anti-neutrino signal observed with Chandler. The prompt and delayed pulses show the number of detected photons as a function of time. Since the decay time constants of the neutrino (EJ-260) and neutron scintillators (EJ-426) employed in this detector are different, the pulse shape discrimination technique can be utilized for discrimination of prompt and delayed signal.

\begin{figure*}
\centering
\includegraphics[width=\linewidth]{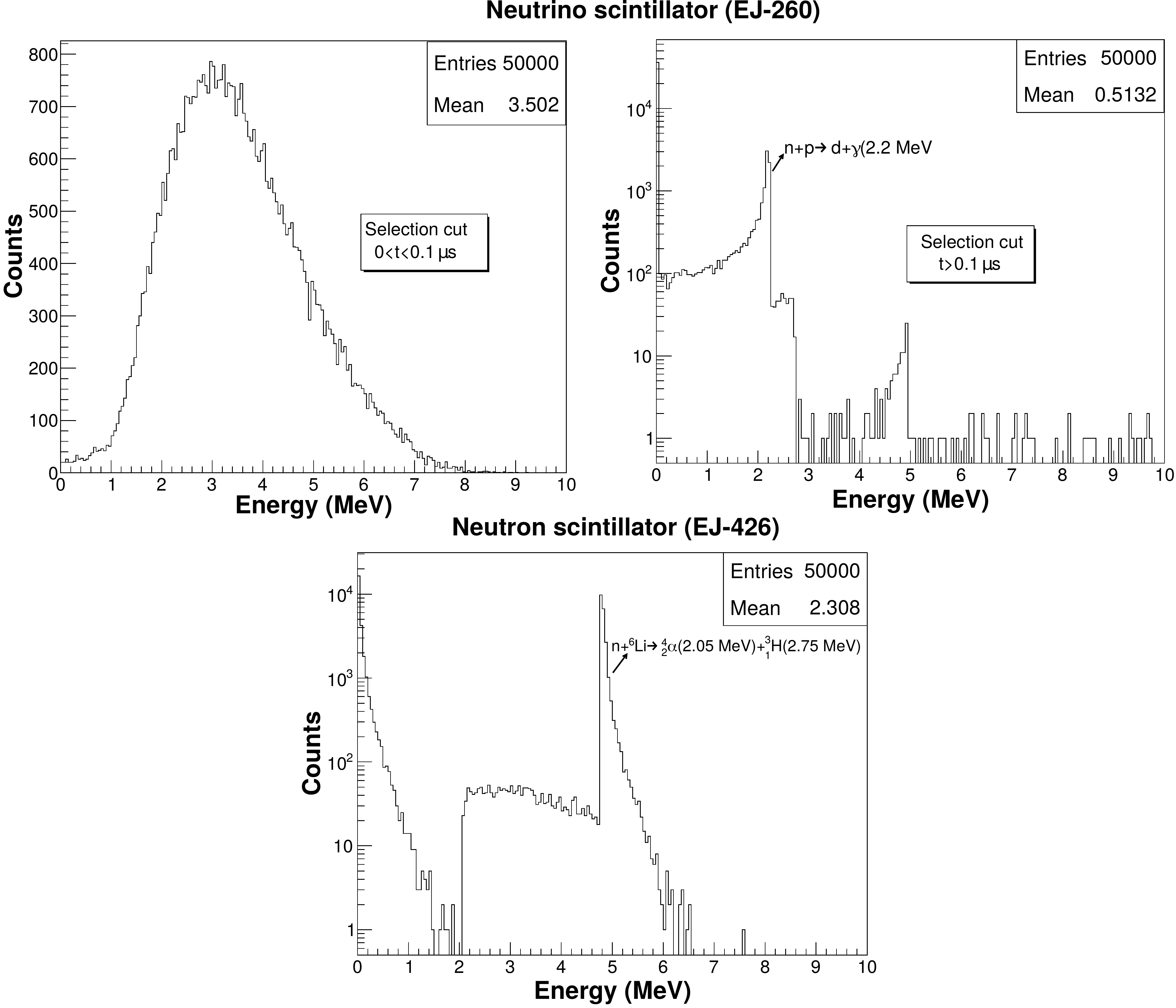}
\caption{Total deposited energy distribution in neutrino scintillators (top) and neutron scintillators (bottom). For the top plots, two different time cuts are applied as calculating total deposited energy in neutrino scintillators.}
\label{fig:fig15}
\end{figure*}

Fig. \ref{fig:fig15} shows the distribution of the total energy deposited in all segments of the Chandler. Since composite detectors employ two distinct scintillator types, the results are shown for both scintillators. The amount of energy accumulated in the neutrino scintillators is calculated by considering the time of energy deposition (before and after 0.1 us for this example).

The macro files that plot these two graphs are also provided for users. To get the same graphs for other detector types, simply change the detector name in the source code of the macros.

\subsection{Comparison metrics}
\label{sec5.2}

When performing an optimization study with NuSD, improving an existing design, or developing a new detector concept, the following physics quantities can be considered as a benchmark for performance evaluation:

\begin{itemize}

\setlength\itemsep{0.em}
\item Intrinsic IBD detection efficiency ($\epsilon _{IBD}$). This is the most significant parameter for an anti-neutrino detector. The following two physics quantities associated with this parameter are also widely used in design optimization and the values of these parameters are usually reported by detector developing communities.
 
\begin{itemize}
\item Neutron capture efficiency. This parameter sets an upper limit on the anti-neutrino detection efficiency. If the neutron escapes from the detector in an IBD event, that event is considered lost. However, the opposite is not true. That is, the neutron can be captured, but the resulting particles may escape from the detector.
\item Neutron capture time. Capturing the neutrons as fast as possible is a desirable feature for all detector types. Reducing the neutron capture time is crucial for improving uncorrelated background rejection efficiency.
\end{itemize}

\item Relative energy resolution. The following equation is used as a metric for optimizing a detector's optical components and comparing the energy resolution performance of different detectors.

\begin{equation}
\label{eqn:last}
\begin{split}
r(E)^{2}  &= ( \frac{ \sigma _{N_e} }{<N_e>} )^{2} 
          +(\frac{\sigma _{LCE}}{<LCE>})^{2} \\
          &+ ( \frac{1}{\sqrt{<N_{p.e}>}} )^{2} 
\end{split}
\end{equation}

To calculate r(E) from the simulation, the following steps are performed: 

\begin{itemize}
\item The G4GenericPrimaryGenerator class is selected as a primary generator by activating the line GENERIC\textunderscore PRIMARY\textunderscore GENERATOR in \textit{NuSD\textunderscore config.h} file. By default, this file generates 1~MeV electron at a random position within the active volume of the detector and fires it in any direction for each event. 

\item During the analysis stage of the simulated data, histograms are made for the emitted photon number ($N_e$), total detected photon number ($N_{p.e}$), and the light collection efficiency (LCE) which is defined as the ratio of the number of photons detected to the total number of photons emitted. The mean values and the standard deviations of these distributions constitute the parameters of Eq. \ref{eqn:last}. The first term in Eq. \ref{eqn:last} corresponds to the intrinsic energy resolution of the scintillator and includes the fluctuation of the light generation process. The second term is defined as the transfer factor and includes the fluctuation of light collection and light detection processes. The last term corresponds to the statistical contribution of the photosensors.

\end{itemize}

\end{itemize}

Table \ref{table:1} comparatively presents the physics quantities defined above for all NuSD detectors. The main purpose of this table is not to interpret the values presented, but to show NuSD users that these physical quantities can be obtained with their own input parameters.

Table \ref{table:2} displays some of the critical input parameters used in the simulation. The findings reported in Table \ref{table:1} strongly depend on these parameters. For example, the LCE is strongly influenced by the absorption length of the scintillator. Similarly, the neutron capture efficiency is strongly dependent on the identity of the nucleus used as the neutron capture agent and its concentration in the active volume. Some input parameters are set to a fixed value at all photon energies. If users wish, they can provide the discrete energy spectrum through the text files provided for them.


\begin{table*}[h]
\caption{Some selected simulation results obtained with NuSD depending on the user input parameters (see Table \ref{table:2}). Neutron capture efficiency strongly depends on the concentration of the neutron capturing elements such as Li and Gd. The light collection efficiency depends on several parameters, the most important of which are shown in Table \ref{table:2}. Here, the quantum efficiency (QE) of the PMT is set to 0.25 and the QE of the SiPM is set to 0.35.}
\begin{tabular}{llllllllll}
\hline
 &
  \multicolumn{3}{l}{\textbf{Inhomogeneous composite}} &
   &
  \multicolumn{3}{l}{\textbf{Homogenous}} &
   &
  \textbf{Inhomogenous} \\ \cline{2-4} \cline{6-8} \cline{10-10} 
                                & SWEANY & CHANDLER & SOLID &  & PROSPECT & NULAT & HSP   &  & PANDA \\ \cline{2-10} 
Time window (us)                & \multicolumn{9}{c}{Neutron capture efficiency (\%)}                \\ \hline
0\textless{}t\textless{}50      & 26     & 21       & 22    &  & 45       & 51    & 68    &  & 38    \\
50\textless{}t\textless{}100    & 11     & 11       & 11    &  & 15       & 14    & 2     &  & 14    \\
100\textless{}t\textless{}150   & 6      & 6        & 6     &  & 5        & 4     & 0     &  & 7     \\
150\textless{}t\textless{}200   & 4      & 4        & 3     &  & 2        & 1     & 0     &  & 3     \\
200\textless{}t\textless{}250   & 2      & 3        & 2     &  & 1        & 0     & 0     &  & 1     \\
250\textless{}t\textless{}500   & 3      & 3        & 2     &  & 0        & 0     & 0     &  & 1     \\ \hline
\begin{tabular}[c]{@{}l@{}}Mean neutron\\ capture time (us)\end{tabular} &
  85 &
  96 &
  80 &
   &
  48 &
  40 &
  16 &
   &
  59 \\ \hline
\begin{tabular}[c]{@{}l@{}}Delayed time \\ window (us)\end{tabular} &
  \multicolumn{9}{c}{\begin{tabular}[c]{@{}c@{}}Anti-neutrino detection efficiency (\%)\\ prompt energy\textgreater{}2 MeV and delayed energy\textgreater{}3 MeV\end{tabular}} \\ \hline
1\textless{}t\textless{}50      & 20     & 16       & 17    &  & 35       & 40    & 33    &  & 24    \\
50\textless{}t\textless{}100    & 9      & 9        & 8     &  & 12       & 11    & 3     &  & 9     \\
100\textless{}t\textless{}150   & 5      & 5        & 5     &  & 4        & 3     & 0     &  & 4     \\
150\textless{}t\textless{}200   & 3      & 3        & 3     &  & 2        & 1     & 0     &  & 2     \\
200\textless{}t\textless{}250   & 2      & 2        & 1     &  & 1        & 0     & 0     &  & 1     \\
250\textless{}t\textless{}500   & 2      & 3        & 2     &  & 0        & 0     & 0     &  & 1     \\
1\textless{}t\textless{}500     & 42     & 38       & 36    &  & 55       & 56    & 37    &  & 42    \\ \hline
\multicolumn{10}{c}{Optical parameters}                                                              \\ \hline
Mean of LCE  (\%)               & 9.18   & 10.56    & 0.51  &  & 3.80     & 3.92 & 7.37  &  & 3.19  \\
Std deviation of LCE            & 1.16   & 0.72     & 0.08  &  & 0.58     & 1.09  & 0.80  &  & 0.18  \\
Intrinsic energy resolution     & 0.01   & 0.01     & 0.01  &  & 0.01     & 0.01  & 0.01  &  & 0.01  \\
Transfer factor                 & 0.13   & 0.07     & 0.16  &  & 0.15     & 0.28  & 0.11  &  & 0.06  \\
Statistical contribution        & 0.03   & 0.03     & 0.13  &  & 0.06     & 0.06  & 0.04  &  & 0.06  \\
Relative energy resolution      & 0.13   & 0.08     & 0.20  &  & 0.16     & 0.29  & 0.12  &  & 0.08  \\
\hline
\end{tabular}
\label{table:1}
\end{table*}



\begin{table*}[]
\caption{There are a large number of input parameters available, but only the most important ones in our judgment are listed here. However, users can run the program using their own experimental input parameters.}
\begin{tabular}{llllllllll}
\hline
\multicolumn{9}{l}{\textbf{Neutron screen material composition \cite{Ely:2013} }} &
   \\  \hline
\begin{tabular}[c]{@{}l@{}}Density = \\ 2.42*g/cm3\end{tabular} &
  \multicolumn{2}{l}{Capture (LiF)} &
  \multicolumn{2}{l}{Scintillation  (ZnS)} &
  \multicolumn{4}{l}{Binder} &
   \\ \hline
Element &
  Enriched Li &
  F &
  Zn &
  S &
  C &
  H &
  O &
  Si &
   \\ \hline
\begin{tabular}[c]{@{}l@{}}Mass fraction\\  (\%)\end{tabular} &
  3.93 &
  12.31 &
  43.58 &
  21.37 &
  6.57 &
  0.17 &
  4.38 &
  7.69 &
   \\ \hline
\multicolumn{9}{l}{\textbf{Custom EJ-309 scintillator material composition \cite{Prospect:2018} }} &
   \\ \hline
\begin{tabular}[c]{@{}l@{}}Density = \\ 0.979*g/cm3\end{tabular} &
  \multicolumn{4}{l}{Homogeneously doped liquid scintillator} &
   &
   &
   &
   &
   \\ \hline
Element &
  Enriched Li &
  H &
  C &
  O &
   &
   &
   &
   &
   \\ \hline
\begin{tabular}[c]{@{}l@{}}Mass fraction\\  (\%)\end{tabular} &
  0.11 &
  9.52 &
  84.14 &
  6.23 &
   &
   &
   &
   &
   \\ \hline
\multicolumn{9}{l}{\textbf{Custom EJ-254 scintillator material composition \cite{Nulat:2015}  }} &
   \\ \hline
\begin{tabular}[c]{@{}l@{}}Density =\\ 1.009*g/cm3\end{tabular} &
  \multicolumn{3}{l}{Homogeneously doped plastic scintillator} &
   &
   &
   &
   &
   &
   \\ \hline
Element &
  Enriched Li &
  H &
  C &
   &
   &
   &
   &
   &
   \\ \hline
\begin{tabular}[c]{@{}l@{}}Mass fraction\\  (\%)\end{tabular} &
  0.11 &
  8.56 &
  91.33 &
   &
   &
   &
   &
   &
   \\ \hline
\multicolumn{10}{l}{\textbf{EJ-335 scintillator material composition}} \\ \hline
\begin{tabular}[c]{@{}l@{}}Density =  \\ 0.89*g/cm3\end{tabular} &
  \multicolumn{3}{l}{Homogeneously doped liquid scintillator} &
   &
   &
   &
   &
   &
   \\ \hline
Element &
  Natural Gd &
  H &
  C &
   &
   &
   &
   &
   &
   \\ \hline
\begin{tabular}[c]{@{}l@{}}Mass fraction\\   (\%)\end{tabular} &
  0.25 &
  11.59 &
  88.16 &
   &
   &
   &
   &
   &
   \\ \hline
\multicolumn{10}{l}{\textbf{Some important optical parameters}} \\ \hline
Material &
  \begin{tabular}[c]{@{}l@{}}Light yield\\ 1/MeV\end{tabular} &
  \begin{tabular}[c]{@{}l@{}}Refractive \\ index\end{tabular} &
  \begin{tabular}[c]{@{}l@{}}Absorption \\ length (cm)\end{tabular} &
   &
   &
   &
   &
   &
   \\ \hline
EJ-200 &
  10000 &
  1.58 &
  300 &
   &
   &
   &
   &
   &
   \\
EJ-260 &
  9200 &
  1.58 &
  \begin{tabular}[c]{@{}l@{}}Energy \\ dependent\end{tabular} &
   &
   &
   &
   &
   &
   \\
EJ-426 &
  40000 &
  1.7 &
  \begin{tabular}[c]{@{}l@{}}Energy\\ dependent\end{tabular} &
   &
   &
   &
   &
   &
   \\
EJ-335 &
  8500 &
  1.49 &
  450 &
   &
   &
   &
   &
   &
   \\
\begin{tabular}[c]{@{}l@{}}Custom \\ EJ-254\end{tabular} &
  7500 &
  1.58 &
  50 &
   &
   &
   &
   &
   &
   \\
\begin{tabular}[c]{@{}l@{}}Custom \\ EJ-309 \cite{Prospect:2018}\end{tabular} &
  8200 &
  1.57 &
  85 &
   &
   &
   &
   &
   &
   \\
Pmma &
  - &
  1.49 &
  300 &
   &
   &
   &
   &
   &
   \\
Air &
  - &
  1 &
  300 &
   &
   &
   &
   &
   &
   \\
   Fep &
  - &
  1.34 &
  300 &
   &
   &
   &
   &
   &
   \\
\begin{tabular}[c]{@{}l@{}}Optical\\ cement\end{tabular} &
  - &
  1.57 &
  300 &
   &
   &
   &
   &
   &
   \\
\begin{tabular}[c]{@{}l@{}}Optical \\ grease\end{tabular} &
  - &
  1.43 &
  300 &
   &
   &
   &
   &
   &
  \\ \hline
\end{tabular}
\label{table:2}
\end{table*}

\section{Conclusions and Future Plans}

In this work, a new simulation framework named NuSD, which is developed in the most general sense to conduct simulation studies in segmented scintillation detectors is introduced. It centers on near-field reactor neutrino detectors developed for the investigation of neutrino oscillation phenomena or potential applications of reactor neutrinos. NuSD combines seven different novel detector concepts developed by various international collaborations under a joint simulation environment, giving the neutrino physics community to explore and test all of these detector technologies.

The benefits of NuSD can be summarized as:

\begin{itemize}
\setlength\itemsep{0.em}
\item Providing a flexible environment to simulate IBD events in seven different detector concepts. It provides a realistic neutrino signal for each detector type and in addition it allows to simulate any particle in different detector types. 

\item Providing a flexible code to perform optimization study for a selected detector type. Users can easily optimize the detector design parameters such as dimensions of each detector component, number of segments, the concentration of neutron capture agents, the thickness of neutron converter material, reflector reflectivity, photo-cathode radius, photo-cathode efficiency. These parameters may vary depending on the chosen detector type.

\item Allowing the users to compare different detectors from various aspects. 

\item Allowing the users to incorporate a new detector concept into NuSD using the supplied classes.

\item Providing many ready-to-use ROOT macro files for the analysis of simulation data.
\end{itemize}

NuSD is a complete simulation framework and a very beneficial tool for physics community, but we're planning to develop its applications in the near-future. So our plan is to have an updated version with the specifications below: 

\begin{itemize}
\setlength\itemsep{0.em}

\item A more user-friendly tool: even if many critical simulation parameters can be managed with user interface commands in the NuSD, but still it can be developed. We're aiming to have an updated version, in which users can control the simulation entirely through user interface commands.  

\item A compatible version with the updated versions of Geant4: after Geant4 10.0.6 version, the implementation method of optical photons has changed considerably and new functionalities have been added. We are planning to adapt these changes quickly in the next version of NuSD.

\item Actively following advances in segmented detector technology: we aim to incorporate innovative detector concepts into the next version of the NuSD. 

\item Extending the scope of the NuSD: it is mainly developed for simulating reactor neutrinos in segmented scintillation detectors but our goal is to extend the NuSD framework to encompass calorimetric studies. 

\item Integrating a detailed signal analysis package into NuSD: it provides some ready-to-use ROOT macro files to analyze raw data but we're planning to build and integrate a detailed signal (physics) analysis package into the next version of the NuSD.

\end{itemize}




\clearpage
\clearpage



\bibliographystyle{elsarticle-num}

\bibliography{sample}

\begin{thebibliography}{10}
\expandafter\ifx\csname url\endcsname\relax
  \def\url#1{\texttt{#1}}\fi
\expandafter\ifx\csname urlprefix\endcsname\relax\def\urlprefix{URL }\fi
\expandafter\ifx\csname href\endcsname\relax
  \def\href#1#2{#2} \def\path#1{#1}\fi

\bibitem{NuTool}
T.~Akindele, et~al., \href{http://dx.doi.org/10.2172/1826602}{Nu tools:
  Exploring practical roles for neutrinos in nuclear energy and security,}\href
  {http://dx.doi.org/10.2172/1826602} {\path{doi:10.2172/1826602}}.
\newline\urlprefix\url{http://dx.doi.org/10.2172/1826602}

\bibitem{Vogel}
P.~Vogel, J.~F. Beacom, Angular distribution of neutron inverse beta decay,
  ${\overline{\ensuremath{\nu}}}_{e}+\stackrel{\ensuremath{\rightarrow}}{p}\rightarrow{e}^{+}+n$,
  Phys. Rev. D 60 (1999) 053003.
\newblock \href {http://dx.doi.org/10.1103/PhysRevD.60.053003}
  {\path{doi:10.1103/PhysRevD.60.053003}}.

\bibitem{WATCHMAN:2015lcq}
M.~Askins, et~al., {The Physics and Nuclear Nonproliferation Goals of WATCHMAN:
  A WAter CHerenkov Monitor for ANtineutrinos}\href
  {http://arxiv.org/abs/1502.01132} {\path{arXiv:1502.01132}}.

\bibitem{Angra}
J.~C. Anjos, et~al., The angra project: Monitoring nuclear reactors with
  antineutrino detectors, AIP Conference Proceedings, 1222~(1) (2010) 427--430.
\newblock \href {http://dx.doi.org/10.1063/1.3399360}
  {\path{doi:10.1063/1.3399360}}.

\bibitem{Fischer:2020htg}
V.~Fischer, E.~Tiras, {Water-based Liquid Scintillator Detector as a New
  Technology Testbed for Neutrino Studies in Turkey}, Nucl. Instrum. Meth. A
  969 (2020) 163931.
\newblock \href {http://arxiv.org/abs/2001.02655} {\path{arXiv:2001.02655}},
  \href {http://dx.doi.org/10.1016/j.nima.2020.163931}
  {\path{doi:10.1016/j.nima.2020.163931}}.

\bibitem{Bat:2021jyq}
A.~Bat, E.~Tiras, V.~Fischer, M.~Kamislioglu, {Low Energy Neutrino Detection
  with a Portable Water-based Liquid Scintillator Detector, }\href
  {http://arxiv.org/abs/2112.03418} {\path{arXiv:2112.03418}}.

\bibitem{Agostinelli}
S.~Agostinelli, J.~Allison, K.~Amako, J.~Apostolakis, et~al., Geant4—a
  simulation toolkit, Nuclear Instruments and Methods in Physics Research
  Section A: Accelerators, Spectrometers, Detectors and Associated Equipment
  506~(3) (2003) 250 -- 303.
\newblock \href {http://dx.doi.org/10.1016/S0168-9002(03)01368-8}
  {\path{doi:10.1016/S0168-9002(03)01368-8}}.

\bibitem{G4Toolkit}
{Geant4 Collaboration}, {Geant4: Book For Toolkit Developers},
  \url{https://geant4-userdoc.web.cern.ch/UsersGuides/ForToolkitDeveloper/fo/BookForToolkitDevelopers.pdf}.

\bibitem{Fowler:2015}
M.~Fowler, UML distilled: a brief guide to the standard object modeling
  language, Addison-Wesley, 2015.

\bibitem{Root}
R.~Brun, F.~Rademakers, Root — an object oriented data analysis framework,
  Nuclear Instruments and Methods in Physics Research Section A: Accelerators,
  Spectrometers, Detectors and Associated Equipment 389~(1) (1997) 81 -- 86.
\newblock \href {http://dx.doi.org/10.1016/S0168-9002(97)00048-X}
  {\path{doi:10.1016/S0168-9002(97)00048-X}}.

\bibitem{G4App}
{Geant4 Collaboration}, {Geant4: Book For Application Developers},
  \url{https://geant4-userdoc.web.cern.ch/UsersGuides/ForApplicationDeveloper/BackupVersions/V10.6c/fo/BookForApplicationDevelopers.pdf}.

\bibitem{QGSP}
A.~Ribon, J.~Apostolakis, A.~Dotti, G.~Folger, V.~Ivanchenko, M.~Kosov,
  V.~Uzhinsky, D.~Wright, Status of geant4 hadronic physics for the simulation
  of lhc experiments at the start of lhc physics program, CERN-LCGAPP 2 (2010)
  2010.

\bibitem{Nulat:2015}
C.~Lane, et~al., {A new type of Neutrino Detector for Sterile Neutrino Search
  at Nuclear Reactors and Nuclear Nonproliferation Applications}\href
  {http://arxiv.org/abs/1501.06935} {\path{arXiv:1501.06935}}.

\bibitem{Nulat:2018}
D.~Xinjian, \href{https://vtechworks.lib.vt.edu/handle/10919/82933}{Development
  and calibration of nulat, a new type of neutrino detector}, Ph.D. thesis
  (2018).
\newline\urlprefix\url{https://vtechworks.lib.vt.edu/handle/10919/82933}

\bibitem{Lens:2007}
C.~Grieb, J.~M. Link, R.~S. Raghavan, Probing active to sterile neutrino
  oscillations in the lens detector, Phys. Rev. D 75 (2007) 093006.
\newblock \href {http://dx.doi.org/10.1103/PhysRevD.75.093006}
  {\path{doi:10.1103/PhysRevD.75.093006}}.

\bibitem{Prospect:2015}
J.~Ashenfelter, et~al., Light collection and pulse-shape discrimination in
  elongated scintillator cells for the {PROSPECT} reactor antineutrino
  experiment, Journal of Instrumentation 10~(11) (2015) P11004--P11004.
\newblock \href {http://dx.doi.org/10.1088/1748-0221/10/11/p11004}
  {\path{doi:10.1088/1748-0221/10/11/p11004}}.

\bibitem{Prospect:2018}
J.~Ashenfelter, et~al., Performance of a segmented 6li-loaded liquid
  scintillator detector for the {PROSPECT} experiment, Journal of
  Instrumentation 13~(06) (2018) P06023--P06023.
\newblock \href {http://dx.doi.org/10.1088/1748-0221/13/06/p06023}
  {\path{doi:10.1088/1748-0221/13/06/p06023}}.

\bibitem{Prospect:2019}
J.~Ashenfelter, et~al., Lithium-loaded liquid scintillator production for the
  {PROSPECT} experiment, Journal of Instrumentation 14~(03) (2019)
  P03026--P03026.
\newblock \href {http://dx.doi.org/10.1088/1748-0221/14/03/p03026}
  {\path{doi:10.1088/1748-0221/14/03/p03026}}.

\bibitem{Prospect:2019:2}
J.~Ashenfelter, et~al., The prospect reactor antineutrino experiment, Nuclear
  Instruments and Methods in Physics Research Section A: Accelerators,
  Spectrometers, Detectors and Associated Equipment 922 (2019) 287--309.
\newblock \href {http://dx.doi.org/https://doi.org/10.1016/j.nima.2018.12.079}
  {\path{doi:https://doi.org/10.1016/j.nima.2018.12.079}}.

\bibitem{Hsp:2020}
M.~Kandemir, A.~Cakir, A reactor antineutrino detector based on hexagonal
  scintillator bars, Nuclear Instruments and Methods in Physics Research
  Section A: Accelerators, Spectrometers, Detectors and Associated Equipment
  953 (2020) 163251.
\newblock \href {http://dx.doi.org/https://doi.org/10.1016/j.nima.2019.163251}
  {\path{doi:https://doi.org/10.1016/j.nima.2019.163251}}.

\bibitem{Panda:2012}
Y.~Kuroda, S.~Oguri, Y.~Kato, R.~Nakata, Y.~Inoue, C.~Ito, M.~Minowa, A mobile
  antineutrino detector with plastic scintillators, Nuclear Instruments and
  Methods in Physics Research Section A: Accelerators, Spectrometers, Detectors
  and Associated Equipment 690 (2012) 41 -- 47.
\newblock \href {http://dx.doi.org/10.1016/j.nima.2012.06.040}
  {\path{doi:10.1016/j.nima.2012.06.040}}.

\bibitem{Panda:2014}
S.~Oguri, Y.~Kuroda, Y.~Kato, R.~Nakata, Y.~Inoue, C.~Ito, M.~Minowa, Reactor
  antineutrino monitoring with a plastic scintillator array as a new safeguards
  method, Nuclear Instruments and Methods in Physics Research Section A:
  Accelerators, Spectrometers, Detectors and Associated Equipment 757 (2014)
  33–39.
\newblock \href {http://dx.doi.org/10.1016/j.nima.2014.04.065}
  {\path{doi:10.1016/j.nima.2014.04.065}}.

\bibitem{Sweany:2015}
M.~Sweany, J.~Brennan, B.~Cabrera-Palmer, S.~Kiff, D.~Reyna, D.~Throckmorton,
  Above-ground antineutrino detection for nuclear reactor monitoring, Nuclear
  Instruments and Methods in Physics Research Section A: Accelerators,
  Spectrometers, Detectors and Associated Equipment 769 (2015) 37--43.
\newblock \href {http://dx.doi.org/https://doi.org/10.1016/j.nima.2014.09.073}
  {\path{doi:https://doi.org/10.1016/j.nima.2014.09.073}}.

\bibitem{Scott:2011}
S.~D. Kiff, N.~Bowden, J.~Lund, D.~Reyna, Neutron detection and identification
  using zns:ag/6lif in segmented antineutrino detectors, Nuclear Instruments
  and Methods in Physics Research Section A: Accelerators, Spectrometers,
  Detectors and Associated Equipment 652~(1) (2011) 412--416, symposium on
  Radiation Measurements and Applications (SORMA) XII 2010.
\newblock \href {http://dx.doi.org/https://doi.org/10.1016/j.nima.2010.07.082}
  {\path{doi:https://doi.org/10.1016/j.nima.2010.07.082}}.

\bibitem{Chandler:2018}
A.~Haghighat, P.~Huber, S.~Li, J.~M. Link, C.~Mariani, J.~Park, T.~Subedi,
  Observation of reactor antineutrinos with a rapidly-deployable surface-level
  detector (2018).
\newblock \href {http://arxiv.org/abs/1812.02163} {\path{arXiv:1812.02163}}.

\bibitem{Chandler:2019}
P.~Huber, J.~M. Link, C.~Mariani, S.~Pal, J.~Park, {CHANDLER} r{\&}d status,
  Journal of Physics: Conference Series 1216 (2019) 012014.
\newblock \href {http://dx.doi.org/10.1088/1742-6596/1216/1/012014}
  {\path{doi:10.1088/1742-6596/1216/1/012014}}.

\bibitem{Chandler:2020}
T.~P. Subedi, {Observation of Reactor Antineutrinos with the CHANDLER
  Detector}, Ph.D. thesis, Virginia Tech. (2020).

\bibitem{Chandler2:2020}
S.~Li, {Detection of Antineutrinos at the North Anna Nuclear Generating
  Station}, Ph.D. thesis, Virginia Tech. (2020).

\bibitem{Solid:2015}
W.~V.~D. Pontseele,
  \href{https://vtechworks.lib.vt.edu/handle/10919/82933}{Characterisation and
  modelling of correlated noise in silicon photomultipliers for the solid
  experiment}, Ph.D. thesis (2017).
\newline\urlprefix\url{https://vtechworks.lib.vt.edu/handle/10919/82933}

\bibitem{Solid:2017}
D.~M. Saunders,
  \href{http://solid-experiment.org/sites/default/files/pdf/thesis/PhD_DanielSaunders.pdf}{First
  data reconstruction and inverse beta decay analysis at the large scale solid
  prototype detector}, Ph.D. thesis (2017).
\newline\urlprefix\url{http://solid-experiment.org/sites/default/files/pdf/thesis/PhD_DanielSaunders.pdf}

\bibitem{Solid2:2017}
Y.~Abreu, et~al., A novel segmented-scintillator antineutrino detector, Journal
  of Instrumentation 12~(04) (2017) P04024--P04024.
\newblock \href {http://dx.doi.org/10.1088/1748-0221/12/04/p04024}
  {\path{doi:10.1088/1748-0221/12/04/p04024}}.

\bibitem{Solid3:2017}
L.~Arnold, W.~Beaumont, D.~Cussans, D.~Newbold, N.~Ryder, A.~Weber, The {SoLid}
  anti-neutrino detectors readout system, Journal of Instrumentation 12~(02)
  (2017) C02012--C02012.
\newblock \href {http://dx.doi.org/10.1088/1748-0221/12/02/c02012}
  {\path{doi:10.1088/1748-0221/12/02/c02012}}.

\bibitem{Solid:2018}
Y.~Abreu, et~al., Performance of a full scale prototype detector at the {BR}2
  reactor for the {SoLid} experiment, Journal of Instrumentation 13~(05) (2018)
  P05005--P05005.
\newblock \href {http://dx.doi.org/10.1088/1748-0221/13/05/p05005}
  {\path{doi:10.1088/1748-0221/13/05/p05005}}.

\bibitem{Solid2:2018}
Y.~Abreu, et~al., Optimisation of the scintillation light collection and
  uniformity for the {SoLid} experiment, Journal of Instrumentation 13~(09)
  (2018) P09005--P09005.
\newblock \href {http://dx.doi.org/10.1088/1748-0221/13/09/p09005}
  {\path{doi:10.1088/1748-0221/13/09/p09005}}.

\bibitem{Ely:2013}
J.~Ely, S.~E.R., S.~M.T., A.~Lintereur., Modeling and simulation optimization
  and feasibility studies for the neutron detection without helium-3 project,
  Tech. rep., Pacific Northwest National Lab. (PNNL) (2013).
\newblock \href {http://dx.doi.org/10.2172/1069205}
  {\path{doi:10.2172/1069205}}.

\end{thebibliography}

\end{document}